\documentclass[aps,pra,amssymb,superscriptaddress,10pt,twocolumn]{revtex4-2}
\usepackage[T1]{fontenc}
\usepackage[latin9]{inputenc}
\usepackage{color}
\usepackage{bm}
\usepackage{amsmath}
\usepackage{amssymb}
\usepackage{graphicx}
\usepackage{subfigure}
\usepackage{setspace}
\usepackage{babel}
\makeatletter

\def\<{\langle}
\def\>{\rangle}

\def\bra#1{\langle #1 |}
\def\ket#1{| #1 \rangle}

\def\ee{\mathrm{e}}

\def\ii{\mathrm{i}}

\makeatother

\begin{document}
\title{Beyond the Rabi model: light interactions with polar atomic systems
in a cavity}
\author{Giovanni Scala}
\affiliation{Dipartimento Interateneo di Fisica, Universit\`a degli Studi di Bari,
I-70126 Bari, Italy}
\affiliation{INFN, Sezione di Bari, I-70125 Bari, Italy}
\affiliation{International Centre for Theory of Quantum Technologies (ICTQT), University of Gdansk, Wita Stwosza 63, 80-308 Gda\'nsk, Poland}
\author{Karolina S\l owik}
\email{karolina@fizyka.umk.pl}

\affiliation{Institute of Physics, Faculty of Physics, Astronomy and Informatics,
Nicolaus Copernicus University in Toru\'{n}, Grudziadzka 5/7, 87-100
Torun, Poland}
\author{Paolo Facchi}
\affiliation{Dipartimento Interateneo di Fisica, Universit\`a degli Studi di Bari,
I-70126 Bari, Italy}
\affiliation{INFN, Sezione di Bari, I-70125 Bari, Italy}
\author{Saverio Pascazio}
\affiliation{Dipartimento Interateneo di Fisica, Universit\`a degli Studi di Bari,
I-70126 Bari, Italy}
\affiliation{INFN, Sezione di Bari, I-70125 Bari, Italy}
\author{Francesco V. Pepe}
\affiliation{Dipartimento Interateneo di Fisica, Universit\`a degli Studi di Bari,
I-70126 Bari, Italy}
\affiliation{INFN, Sezione di Bari, I-70125 Bari, Italy}
\date{\today}

\begin{abstract}
The Rabi Hamiltonian, describing the interaction between
a two-level atomic system and a single cavity mode of the electromagnetic field, is one of the fundamental models in quantum optics. The model becomes exactly solvable by considering an atom without permanent dipole moments, whose excitation energy is quasi-resonant with the cavity photon energy, and by neglecting the non resonant (counter-rotating) terms. In this case, after including the decay of either the atom or the cavity mode to a continuum, one is able to derive the well-known phenomenology of quasi-resonant transitions, including the fluorescence triplets. In this work we consider the most general Rabi model, incorporating the  effects of permanent atomic electric dipole moments, and, based on a perturbative analysis, we compare the intensities of emission lines induced by rotating terms, counter-rotating terms and parity-symmetry-breaking terms. The analysis reveals that the emission
strength related to the existence of permanent dipoles may surpass the one due to the counter-rotating interaction terms, but is usually much weaker than the emission due to the main, resonant coupling. This ratio can be modified in systems with a reduced dimensionality or by engineering the energy spectral  density of the continuum. 
\end{abstract}
\pacs{....}
\maketitle

\section{\label{sec:level1}Introduction}

The Rabi model is a fundamental tool in quantum optics. It describes
the coupling of a two-level system and a bosonic field mode~\cite{xie2017}, extending beyond the simpler Jaynes-Cummings
interaction, in which the creation of a photon is always accompanied by annihilation
of the atomic excitation and vice versa~\cite{shore1993}. The Rabi model
additionally accounts for the less intuitive processes of pairwise
creation or annihilation of excitations in the atomic and photonic
subsystems. The probability of these processes grows with the light-matter
coupling constant and becomes significant in the so-called ultrastrong
coupling regime, in which the coupling constant becomes comparable
to the energy of the system~\cite{forn2019}. Numerous experimental realizations
include superconducting systems~\cite{bourassa2009,niemczyk2010},
quantum wells~\cite{gunter2009, zhang2016}, photonic waveguide arrays
\cite{crespi2012}, molecular ensembles~\cite{george2016}, cold atoms
\cite{schneeweiss2018}, etc. In all these systems the extension beyond
the Jaynes-Cummings interaction may lead to considerably different physics: in particular, to a ground state with
a nonvanishing number of excitations, squeezing dynamic, and a significant
modification of the spectra~\cite{braak2011, chen2012, xie2017}.
Remarkably, analytical solutions of the Rabi model have been developed
only in the last decade~\cite{braak2011, chen2012}. 

\begin{figure}
\centering
\includegraphics[width=0.4\textwidth]{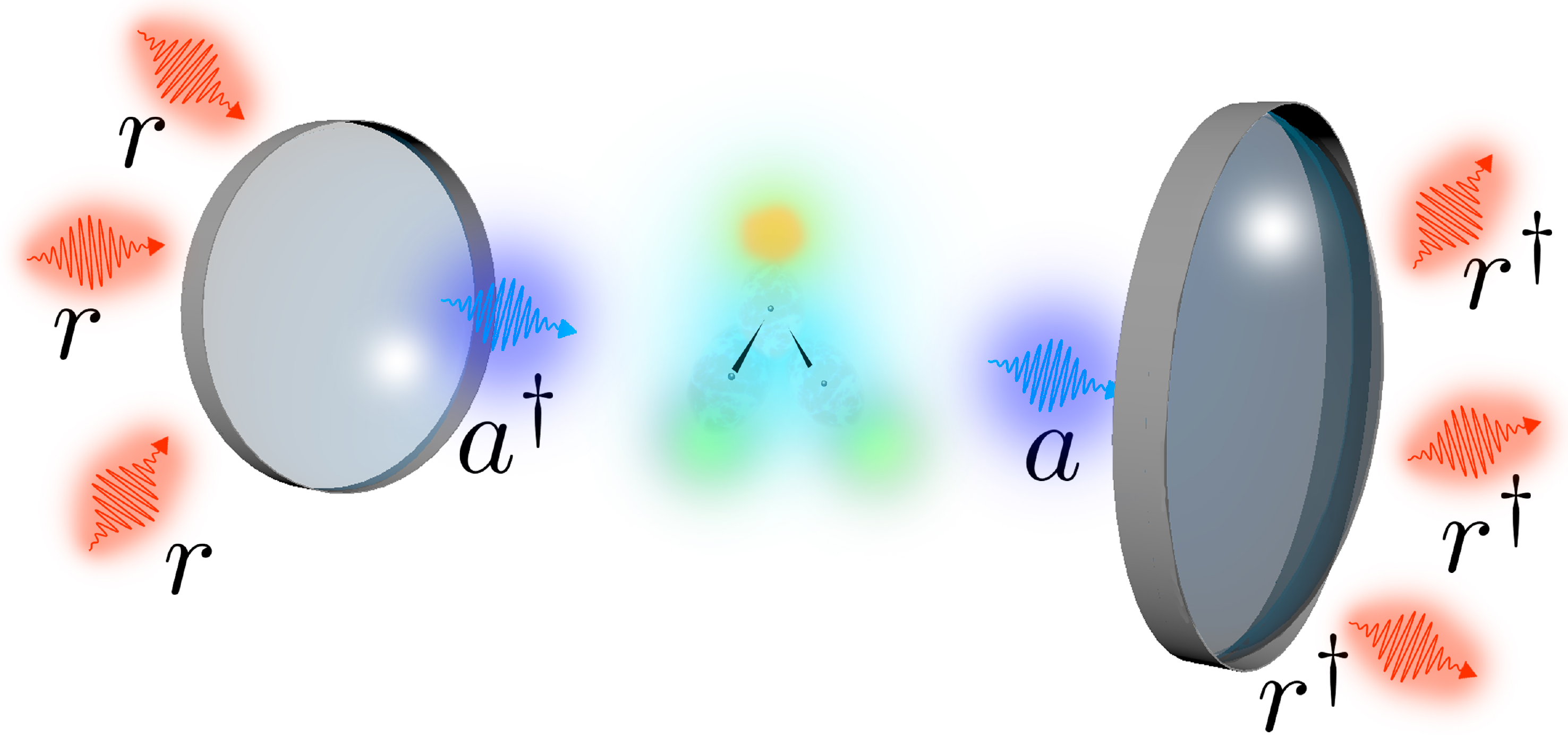}
\caption{Sketch of the system under study: a two-level polar atomic system in a lossy cavity represented by two semitransparent mirrors. The non-uniform charge distribution is shown in green for a higher concentration of positive charges and in orange for the negative charges. The annihilation and creation operators of the electromagnetic modes are denoted as $a$, $a^\dagger$ for the cavity (blue), and $r$, $r^\dagger$ for the reservoir (red). }
\label{fig:Hex}
\end{figure}

The Rabi model describes light-matter interaction,
where the electromagnetic field induces transitions between the eigenstates
of a two-level atomic system. A particular mechanism is related
to a coupling of the electromagnetic radiation with a transition dipole
moment element induced between a pair of atomic eigenstates. However,
simple two-level systems may display versatile physical features, beyond the
traditional Rabi model: a particular example is a coupling scenario
where the electromagnetic field introduces energy shifts of the eigenstates
rather than transitions between them~\cite{kibis2009,paspalakis2013}. A simple realization exploits
atomic systems with permanent dipole moments, such as polar molecules
or asymmetric quantum dots. Due to the interplay of permanent and
induced electric dipole moments, polar systems are a playground where
a richer physics of light-matter interactions can be realized: polar quantum
systems have been proposed for THz radiation sources~\cite{kibis2009} based on quantum dots~\cite{chestnov2017}
or molecular ensembles~\cite{gladysz2020}. They can be exploited for
squeezed light generation~\cite{koppenhofer2016,anton2017} and they support nonlinear optical absorption~\cite{paspalakis2013}. Recently, the impact of spatial asymmetry
of a quantum system on its spontaneous emission properties has been
investigated~\cite{scala2020}. 

The aforementioned works are among the plethora of possibilities provided by asymmetric
quantum systems that simultaneously support light-matter interactions
through three types of terms. These include the Jaynes-Cummings
terms and the counter-rotating terms, both involving transition
dipole moments of the atomic system. The third type of terms involves permanent dipoles, i.e.\ nonvanishing expectation values of the dipole moment operator in the atomic eigenstates. For the numerous applications listed above it is essential
to identify conditions in terms of experimentally tunable model parameters
where different contributions significantly influence the
system's optical response. The aim of this work is to study
the relative impact of these three contributions, and demonstrate with
simple examples the possibility of performing density-of-states
engineering. Our analysis follows the methodology introduced in Ref.~\cite{savenko2012}, but extends it to include
all the three interaction mechanisms. 

The paper is organized as follows: a two-level atomic system without
inversion symmetry, coupled to a single-mode electromagnetic field,
is introduced in Section II. Next, we apply a perturbative framework
to find a ladder of eigenstates and the correponding energies in Section
III. Transitions between these eigenstates upon a coupling with an
external lossy cavity are described in Section IV, which ends the
analytical part. Numerical  examples of  systems with
low to moderate light-matter coupling strengths are given in Section
V. In Appendix~\ref{app:test}, we discuss the validity of the perturbative approach, while details of calculations of the spectral distribution of emitted photons are given in Appendix~\ref{app:resolvent}. 

\section{\label{sec:level2}Hamiltonian of the system}
Let us consider a two-level system with a ground and excited state
denoted respectively as $|g\rangle$, $|e\rangle$, separated by the excitation energy $\hbar\omega_{a}$. The system is described by the set of Pauli
operators 
\begin{align}
\sigma_{-} & =|g\rangle\langle e|, \quad \sigma_{+}=|e\rangle\langle g|, \\
\sigma_{z} & =|e\rangle\langle e|-|g\rangle\langle g| .
\end{align}
This system interacts with a single electromagnetic cavity mode, represented by the field operators $a$ and $a^{\dagger}$, satisfying the canonical commutation algebra
\begin{equation}
[a,a^{\dagger}]=1, \quad [a,a] = [a^{\dagger},a^{\dagger}] = 0.
\end{equation}
The Hamiltonian $H$ of the coupled system can be divided
in two parts 
\begin{equation}\label{key-1}
H = H_{\mathrm{JC}} + V ,
\end{equation}
with the first term
\begin{equation}\label{eq:HJC}
H_{\mathrm{JC}}=\hbar\omega_c a^{\dagger}a + \frac{\hbar\omega_a}{2} \sigma_{z} + \hbar g_R(\sigma_{+}a+a^{\dagger}\sigma_{-}) ,
\end{equation}
known as  the Jaynes-Cummings (JC) Hamiltonian~\cite{shore1993}, that describes quasi-resonant transitions between the atomic excitations and photons.
Here, $g_{R}$ is the coupling strength of the resonant JC term. The results in the following analysis are independent of the coupling mechanism and the specific expressions of the coupling constants in terms of microscopic parameters. In the case of an atom coupled to one mode of a 3D rectangular cavity, the coupling constant reads \mbox{$g_{R}=-\bm{d}_{\mathrm{eg}}\cdot\bm{\epsilon}\sqrt{\hbar\omega_{c}/2\epsilon_{0}\mathcal{V}}$}, where $\bm{d}_{\mathrm{eg}}=\langle e|\bm{d}|g\rangle$ represents the off-diagonal matrix element of the electric dipole operator $\boldsymbol{d}$ of the atom, $\boldsymbol{\epsilon}$ is
the polarization vector of the cavity mode, $\epsilon_{0}$ the vacuum electric permittivity
and $\mathcal{V}$ the cavity volume. 

The ``perturbation'' term $V$ in Eq.~\eqref{key-1} accounts for all the terms that are not represented in the exactly solvable Jaynes-Cummings Hamiltonian, namely the counter-rotating
(CR) transitions between atom and cavity excitations and the terms proportional to the diagonal matrix elements of the atomic dipole moment:  
\begin{align}
V & =H_{\mathrm{CR}}+H_{\mathrm{AS}},\label{eq:HS_HR}\\
H_{\mathrm{CR}} & =\hbar g_{R}\left(\sigma_{+}a^{\dagger}+\sigma_{-}a\right),\\
H_{\mathrm{AS}} & =\hbar \left[ g_{S}\left(\sigma_{z}+\boldsymbol{1}\right) + g'_{S}\left(\sigma_{z}-\boldsymbol{1}\right) \right] \left(a+a^{\dagger}\right).
\end{align}
with 
\begin{align}
g_{S} & =-\bm{d}_{\mathrm{ee}}\cdot\bm{\epsilon}\sqrt{\hbar\omega_{c}/8\epsilon_{0}\mathcal{V}} \\
g'_{S} & =-\bm{d}_{\mathrm{gg}}\cdot\bm{\epsilon}\sqrt{\hbar\omega_{c}/8\epsilon_{0}\mathcal{V}}
\end{align}
proportional to the expectation values of the atomic dipole moment on the excited and ground state, respectively. In this article, we will focus for definiteness on the case $g'_S=0$.

Note that the expectation value of a dipole moment operator described
only by off-diagonal elements $\boldsymbol{d}_{\mathrm{eg}}|e\rangle\langle g|+\boldsymbol{d}_{\mathrm{eg}}^*|g\rangle\langle e|$
may be nonzero only in presence of transitions between the eigenstates
that may be induced with the external electric field. Therefore, these
elements correspond to induced transition dipoles. On the other hand, the diagonal
element describes the permanent dipole moment of the excited state.
Notably, permanent dipole moments are sustained by polar systems,
i.e.\ systems without inversion symmetry~\cite{kibis2009}.
For this reason we will refer to the last Hamiltonian term as the
``asymmetry term'' or ``diagonal term'' and mark it with the AS subscript. Finally, note
that while the Hamiltonian $H_{\mathrm{JC}}$ preserves the number
of excitations, $H_{\mathrm{AS}}$ ($H_{\mathrm{CR}}$) describes
a modification of this number by 1 (respectively 2). 

\section{Perturbative analysis}

In the following analysis we will treat $V$ as a perturbation with
respect to the Hamiltonian $H_{\mathrm{JC}}$. The eigenvalues
of $H_{\mathrm{JC}}$ correspond to the $0$th order perturbation term 
\begin{equation}
E_{n}^{s(0)}=\hbar\omega_{c}\left(n-\frac{1}{2}\right)+s\hbar\sqrt{\frac{\left(\omega_{c}-\omega_{a}\right)^{2}}{4}+ng_{R}^{2}}\label{energieslevel}
\end{equation}
for $n=0,1,\dots$ and $s=\pm1$, and the eigenstates are 
\begin{equation}
\left|n_{s}^{\left(0\right)}\right\rangle =A_{n}^{s}\left|g;n\right\rangle +B_{n}^{s}\left|e,n-1\right\rangle ,\label{eq:psi0}
\end{equation}
with 
\begin{align}
A_{n}^{s} & =\frac{E_{n}^{s(0)}-\hbar\omega_{c}\left(n-1\right)-\hbar\omega_{a}/2}{\sqrt{\Bigl(E_{n}^{s(0)}-\hbar\omega_{c}(n-1)-\hbar\omega_{a}/2\Bigr)^{2}+\hbar^{2}g_{R}^{2}n}},\\
B_{n}^{s} & =\frac{\hbar g_{r}\sqrt{n}}{\sqrt{\Bigl(E_{n}^{s(0)}-\hbar\omega_{c}(n-1)-\hbar\omega_{a}/2\Bigr)^{2}+\hbar^{2}g_{R}^{2}n}}.
\end{align}
The pair $\left\{ \left|n_{s}^{\left(0\right)}\right\rangle \right\} _{s=\pm}$
defines a two-dimensional manifold $\mathcal{E}_{\mathrm{JC}}\left(n\right)$, which
is the set of states with a fixed number of excitations $n$ (see
Fig.~\ref{fig:transitions}). We denote it with the JC subscript,
since the notion of manifold will be generalized in the perturbed
picture.

In the perturbation Hamiltonian $V$, the counter-rotating term $H_{\mathrm{CR}}$
is described by the same coupling constant $g_{R}$ as the interaction
term of the unperturbed Hamiltonian $H_{\mathrm{JC}}$. However, the
transition rates due to $H_{\mathrm{CR}}$ are much smaller far from the ultrastrong coupling regime $g_{R}\ll\omega$.
Therefore, perturbation theory is justified up to
moderate coupling strengths (see Appendix~\ref{app:test} for quantitative details).
We characterize the modified eigenstates of the time-independent
perturbation theory up to second order, with the wavefunction
expansion given by 
\begin{equation}
\left|n_{s}\right\rangle =\left|n_{s}^{\left(0\right)}\right\rangle +\left|n_{s}^{\left(1\right)}\right\rangle +\left|n_{s}^{\left(2\right)}\right\rangle .
\end{equation}
The first-order correction reads 
\begin{equation}
\left|n_{s}^{\left(1\right)}\right\rangle =\sum_{m\neq n}\sum_{\alpha=\pm}\frac{V_{mn}^{\alpha s}}{E_{nm}^{s\alpha}}\left|m_{\alpha}^{\left(0\right)}\right\rangle 
,\label{def:first_order_state}
\end{equation}
where $E_{nm}^{s\alpha}=E_{n}^{s\left(0\right)}-E_{m}^{\alpha\left(0\right)}$
and $V_{mn}^{\alpha s}=\left\langle m_{\alpha}^{\left(0\right)}|V|n_{s}^{\left(0\right)}\right\rangle $, namely
\begin{align}
V_{mn}^{\alpha s}= & \hbar g_{R}\Bigg(\sqrt{n-1}B_{n}^{s}A_{n-2}^{\alpha}\delta_{m,n-2}\nonumber \\
 & +\sqrt{n+1}A_{n}^{s}B_{n+2}^{\alpha}\delta_{m,n+2}\Bigg)\nonumber \\
+ & 2\hbar g_{S}B_{m}^{\alpha}B_{n}^{s}\left(\sqrt{n-1}\delta_{m,n-1}+\sqrt{n}\delta_{m,n+1}\right).\label{eq:Vmn}
\end{align}
The above equation shows that the perturbed eigenstates include states with $m=n\pm1$ coupled by $g_{S}$ and states with $m=n\pm2$
coupled by $g_{R}$, which follows directly from the
$H_{\mathrm{AS}}$ and $H_{\mathrm{CR}}$ Hamiltonians. The inclusion of
the second-order correction leads to
\begin{align}\label{eq:ket_2nd_order}
\left|n_{s}\right\rangle  & =\left(1-\frac{1}{2}\sum_{k}\sum_{\alpha=\pm}\left(\frac{V_{nk}^{s\alpha}}{E_{nk}^{s\alpha}}\right)^{2}\right)\left|n_{s}^{\left(0\right)}\right\rangle \\
 & +\sum_{k}\sum_{\alpha=\pm}\left(\frac{V_{kn}^{\alpha s}}{E_{nk}^{s\alpha}}+\sum_{l}\sum_{\beta=\pm}\frac{V_{kl}^{\alpha\beta}V_{ln}^{\beta s}}{E_{nk}^{s\alpha}E_{nl}^{s\beta}}\right)\left|k_{\alpha}^{\left(0\right)}\right\rangle. \nonumber 
\end{align}
Based on the above result, we define the generalized (but always two-dimensional) manifolds
$\mathcal{E}\left(n\right)=\left\{ \left|n_{s}\right\rangle \right\} _{s=\pm}$.
According to second order perturbation, the eigenstate $\left|n_{s}\right\rangle $
includes contributions with different numbers of excitations
$\left\{ n,n\pm1,\dots,n\pm4\right\} $, with the label $n$ referring
to the central component, which yields by far the leading contribution for weak enough coupling strengths $g_{R,S}$, for which
the theory is applicable. 

The correction $V$ does not perturb the eigenvalues at the first
order, because $V_{nn}^{s\sigma}=\left\langle n_{s}|V|n_{\sigma}\right\rangle =0$.
At the second order, the energy eigenvalues are $E_{n}^{s}=E_{n}^{s\left(0\right)}+E_{n}^{s\left(1\right)}+E_{n}^{s\left(2\right)}$,
with $E_{n}^{s\left(1\right)}=0$ and 
\begin{equation}
E_{n}^{s\left(2\right)}=\sum_{k\neq n}\sum_{\iota=\pm}\frac{\left(V_{kn}^{\iota s}\right)^{2}}{E_{nk}^{s\iota}}.\label{eq:energy_2nd_order}
\end{equation}

\section{outcoupling}

In this section, we assume the cavity mirrors to be semi-transparent,
so that the cavity mode described by $a$ and $a^{\dagger}$ may exchange photons with
an external reservoir: 
\begin{equation}\label{key-2}
H_{\mathrm{ext}}=\hbar\sqrt{\frac{\Gamma}{2\pi}} \int d\omega \sqrt{\mathcal{P}(\omega)} \left(a r^{\dagger}(\omega)+a^{\dagger} r(\omega)\right),
\end{equation}
where the operators $r(\omega)$ and $r^{\dagger}(\omega)$ are related to orthogonal reservoir modes with energy $\hbar\omega$, and $\mathcal{P}(\omega)$ is a form factor that takes into account both the density of states and the energy dependence of the coupling, with \mbox{$\mathcal{P}(\omega_c)=1$} for convenience. The constants are fixed in such a way that $\Gamma$ coincides with the perturbative decay rate of a single cavity photon towards the continuum,
\begin{equation}
\Gamma_{1\to 0} = \frac{2\pi}{\hbar^2} \int d\omega |\bra{0;\omega_R} H_{\mathrm{ext}} \ket{1;0_R} |^2 \delta(\omega - \omega_c) = \Gamma, 
\end{equation}
with $\ket{0_R}$ standing for the reservoir vacuum, annihilated by all the $r$ operators, and $\ket{\omega_R}= r^{\dagger}(\omega)\ket{0_R}$ being a generic single-photon state with a given energy $\omega$, while the transition rate from the $n$-photon to the $(n-1)$-photon state of the cavity reads $\Gamma_{n\to n-1}=n\Gamma$.

Here, we will compute through the Fermi golden rule the decay rate and final photon energy distribution of the dressed atom-cavity states found in the previous section. In the perturbative regime, the transition from an initial state $\ket{n_s}$ to a final state $\ket{n'_{s'}}$, as defined in Eq.~(\ref{eq:ket_2nd_order}), corresponds to the transition frequency 
\begin{equation}
\omega_{nn'}^{ss'}= \frac{E_n^s - E_{n'}^{s'}}{\hbar}
\end{equation}
and is determined by the matrix elements
\begin{equation}\label{element}
\bra{n'_{s'};\omega_R} H_{\mathrm{ext}} \ket{n_s;0_R} = \hbar \sqrt{\frac{\Gamma}{2\pi} \mathcal{P}(\omega)} \, \bra{n'_{s'}} a \ket{n_s} ,
\end{equation}
which are evaluated on-shell in the expression of the specific decay rate towards channel $n'_{s'}$
\begin{equation}\label{Gamma}
\Gamma_{nn'}^{ss'} = \Gamma \left| \bra{n'_{s'}} a \ket{n_s} \right|^2 \mathcal{P} ( \omega_{nn'}^{ss'} ) ,
\end{equation}
that contribute to the total decay rate of the initial state:
\begin{equation}\label{Gammatot}
\Gamma_{n,s} = \sum_{n',s'} \Gamma_{nn'}^{ss'} = \Gamma \sum_{n',s'} \left| \bra{n'_{s'}} a \ket{n_s} \right|^2 \mathcal{P} ( \omega_{nn'}^{ss'} ) .
\end{equation}
Notice that i) the above expressions are valid provided that all the channels are characterized by different transition energies, otherwise interference effects occur, and ii) the form factor $\mathcal{P}$ must vanish for $\omega$ below the threshold for photon emission. The specific and total decay rates also appear in the frequency distribution of the final photons, derived in Appendix~\ref{app:resolvent},
\begin{align}\label{distribution}
S_{n,s}(\omega) & = \sum_{n',s'} S_{nn'}^{ss'}(\omega) \nonumber
\\ & = \frac{\Gamma}{2\pi} \sum_{n',s'} \frac{\left| \bra{n'_{s'}} a \ket{n_s} \right|^2 \mathcal{P} ( \omega_{nn'}^{ss'} )}{ (\omega- \omega_{nn'}^{ss'} - \Delta_{n,s})^2 + \Gamma_{n,s}^2/4  },
\end{align}
where $S_{nn'}^{ss'}(\omega)$ are specific spectral distributions, related to a single decay channel. The spectral distribution is characterized by the presence of Lorentzian peaks around the transition frequencies, all shifted by 
\begin{equation}\label{Delta}
\Delta_{n,s} = \frac{\Gamma}{2\pi} \sum_{n',s'} \left| \bra{n'_{s'}} a \ket{n_s} \right|^2 \mathrm{P} \int d\omega \frac{ \mathcal{P} ( \omega )}{\omega - \omega_{nn'}^{ss'}} ,
\end{equation}
with $\mathrm{P}\int$ denoting principal value integration. The specific decay rates also determine the weight of each channel. Therefore, the relevance of one channel compared to another one can crucially depend on the transition energy, through the form factor:
\begin{equation}\label{weights}
\frac{\Gamma_{nn'}^{ss'}}{\Gamma_{nn''}^{ss''}} = \frac{\left| \bra{n'_{s'}} a \ket{n_s} \right|^2}{\left| \bra{n''_{s''}} a \ket{n_s} \right|^2} \frac{\mathcal{P} ( \omega_{nn'}^{ss'} )}{\mathcal{P} ( \omega_{nn''}^{ss''} )} . 
\end{equation}
For example, the form factor can be characterized by a power-law behavior for the transition frequencies, $\mathcal{P}(\omega)\sim \omega^p$, as it occurs for free-space photons: in this case, lower-energy channels can be heavily hindered in favor of the higher-energy ones, despite being characterized by a larger matrix element of the operator $a$ in Eq.~(\ref{weights}). On the other hand, the relevance of a channel can be enhanced by engineering the continuum in order to obtain a form factor peaked around the frequency of interest: this can be done by coherently coupling the cavity mode with a single mode of a second cavity, broadened by losses towards free space. 

We now evaluate the matrix element $a$, appearing in Eqs.~\eqref{element}--\eqref{weights}. At the
 $0$-th order, we find directly from Eq.~(\ref{eq:psi0}) 
\begin{multline}\label{eq:Frl}
\left\langle {n'}_{s'}^{\left(0\right)}|a |n_{s}^{\left(0\right)}\right\rangle  =  c_{n}^{s's}\delta_{n',n-1}  \\
 =  \left(\sqrt{n}A_{n}^{s}A_{n-1}^{s'}+\sqrt{n-1}B_{n}^{s}B_{n-1}^{s'}\right)\delta_{n',n-1}.
\end{multline}
This equation shows that only transitions between two ``adjacent'' 
manifolds $\mathcal{E_{\mathrm{JC}}}\left(n\right)$ and $\mathcal{E}_{\mathrm{JC}}\left(n-1\right)$
are allowed in the Jaynes-Cummings model, as expected. Our goal is
to analyze the emission probability via Eq.~\eqref{distribution}
when the eigenstates are corrected by the perturbation term $V$ in
Eq.~(\ref{eq:HS_HR}) and the spectrum is defined in Eq.~(\ref{energieslevel}).
In the perturbed expression for $\left\langle {n'}_{s'}|a|n_s\right\rangle $,
obtained from the eigenstates $\left|n_{s}\right\rangle$ in Eq.~\eqref{eq:ket_2nd_order}, we consider
all the correction terms up to the second order in the coupling strengths
$g_{R}$ and $g_{S}$. 
Using $V_{nn}^{ss'}=0$, we find
\begin{widetext}
\begin{align}\left|\left\langle {n'}_{s'}|a|n_s \right\rangle \right|^{2}= & \left|\sum_{\alpha=\pm}\left(\frac{V_{n'+1,n}^{\alpha s } }{E_{n,n'+1}^{s \alpha}}c_{n'+1}^{s'\alpha}+\frac{V_{n+1,n'}^{\alpha s'}}{E_{n',n+1}^{s'\alpha}}c_{n}^{\alpha s } \right)\right|^{2}+\delta_{n',n-1}\left\{\left[1-\sum_{k}\sum_{\alpha=\pm} \left( \left(\frac{V_{nk}^{s\alpha}}{E_{nk}^{s\alpha}}\right)^{2} + \left(\frac{V_{n'k}^{s'\alpha}}{E_{n'k}^{s'\alpha}}\right)^{2} \right) \right]\left|c_{n}^{s's}\right|^{2}\right.\nonumber\\
 & \left.+2c_{n}^{s's}\sum_{k}\sum_{\alpha,\beta=\pm}\left(\frac{V_{nk}^{\alpha\beta}V_{kn}^{\beta s}}{E_{nn}^{s\alpha}E_{nk}^{s\beta}}c_{n}^{s'\alpha}+\frac{V_{k-1,n-1}^{\alpha s'}}{E_{n-1,k-1}^{s'\alpha}}\frac{V_{kn}^{\beta s} }{E_{nk}^{s\beta}}c_{k}^{\alpha\beta}+\frac{V_{n-1,k}^{\alpha\beta}V_{k,n-1}^{\beta s'}}{E_{n-1,n-1}^{s'\alpha}E_{n-1,k}^{s'\beta}}c_{n}^{\alpha s}\right)\right\},
\end{align}
\end{widetext}
where $c_n^{ss^\prime}$ is given in Eq.~(\ref{eq:Frl}). 
Inclusion of the perturbation Hamiltonian allows transitions to new
manifolds. Close inspection of the form of the perturbation $V$ in
Eq.~(\ref{eq:Vmn}), combined with the above expression for $\left|\left\langle {n'}_{s'}|a|n_s\right\rangle \right|^{2}$,
reveals that transitions between $\mathcal{E}\left(n\right)\to\mathcal{E}\left(n'=n\right)$
and $\mathcal{E}\left(n\right)\to\mathcal{E}\left(n'=n-2\right)$
are due to the {diagonal-coupling Hamiltonian $H_{\mathrm{AS}}$}
and the transition between $\mathcal{E}\left(n\right)\to\mathcal{E}\left(n-3\right)$
are due to $H_{\mathrm{CR}}$. The latter would also give rise to
$\mathcal{E}\left(n\right)\to\mathcal{E}\left(n+1\right)$
transitions, which are suppressed in the reservoir vacuum state $|0_R\rangle$.
One might argue that closer manifolds are favorite, as JC transitions involve only adjacent manifolds, yielding a stronger contribution 
from the diagonal coupling rather than the counter-rotating terms. On the other hand, transitions at higher frequencies
can contribute with a higher intensity due to the larger form factor $\mathcal{P}$. Therefore, a quantitative comparison is needed
to evaluate the spectrum and properly characterize the behavior in different regimes. 

\section{Results}

\begin{figure}
\includegraphics[width=0.75\linewidth]{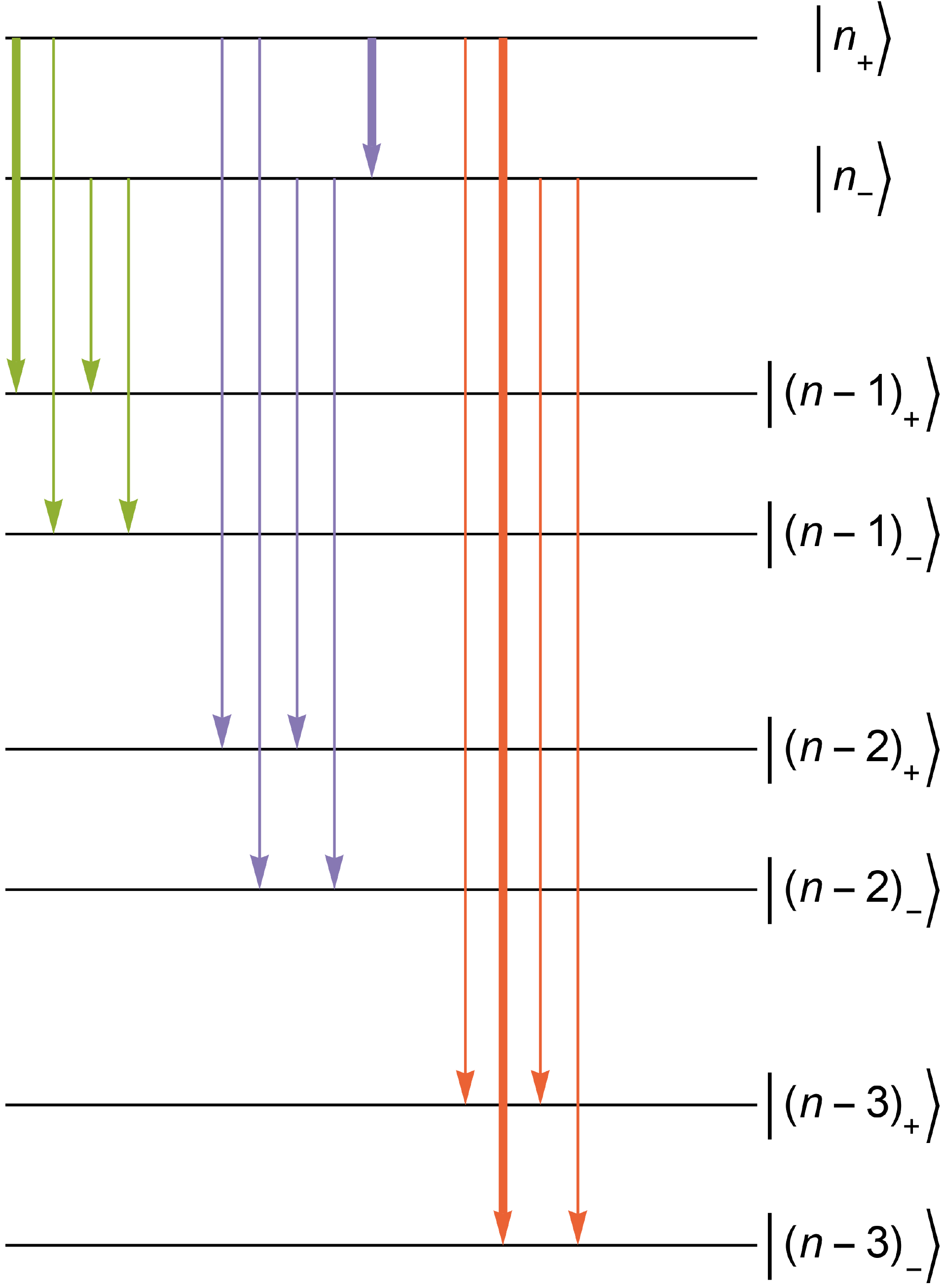}
\caption{The first set of transitions (green) is allowed by $H_{\mathrm{JC}}$, as expressed
in Eq.~\eqref{eq:Frl}. Five new transition lines (violet) towards $n'=n$ and $n'=n-2$ are origin at $H_{\mathrm{AS}}$. Transitions
due to the counter-rotating Hamiltonian $H_{\mathrm{CR}}$ connect manifold $\mathcal{E}(n)$ to $\mathcal{E}(n-3)$.
The three thicker lines are studied in more details in Fig.~\ref{fig:amplitudes}.}
\label{fig:transitions}
\end{figure}

In this section we analyze the transitions shown in Fig. \ref{fig:transitions},
with a special emphasis on the ones induced by the perturbation Hamiltonian
$H_{\mathrm{AS}}$ and $H_{\mathrm{CR}}$. We will investigate relative
emission strengths of transitions that origin from different Hamiltonian
contributions as functions of the coupling constants $g_{R,S}$.

Figure \ref{fig:transitions} depicts the thirteen allowed transitions
from a given manifold $\mathcal{E}\left(n\right)$, connecting respectively manifolds
$\mathcal{E}\left(n\right)\rightarrow\mathcal{E}\left(n-1\right)$
(JC interaction term, green arrows), $\mathcal{E}\left(n\right)\rightarrow\mathcal{E}\left(n\right)$
and $\mathcal{E}\left(n\right)\rightarrow\mathcal{E}\left(n-2\right)$
 (AS Hamiltonian, purple arrows), $\mathcal{E}\left(n\right)\rightarrow\mathcal{E}\left(n-3\right)$
(CR contribution, red arrows). 
\begin{figure*}
\subfigure[resonant]{\includegraphics[width=0.48\textwidth]{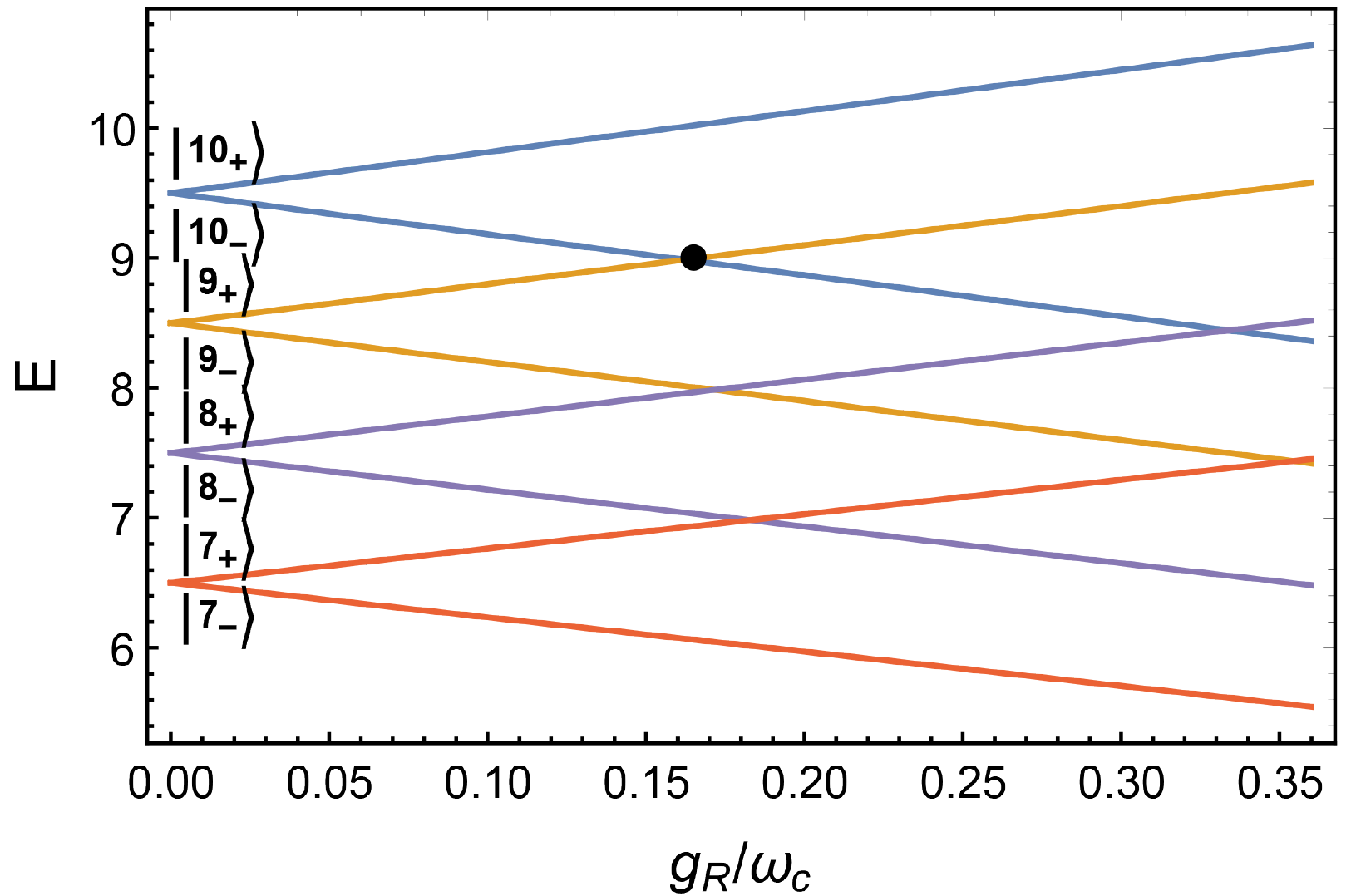}}
\subfigure[detuned]{\includegraphics[width=0.48\textwidth]{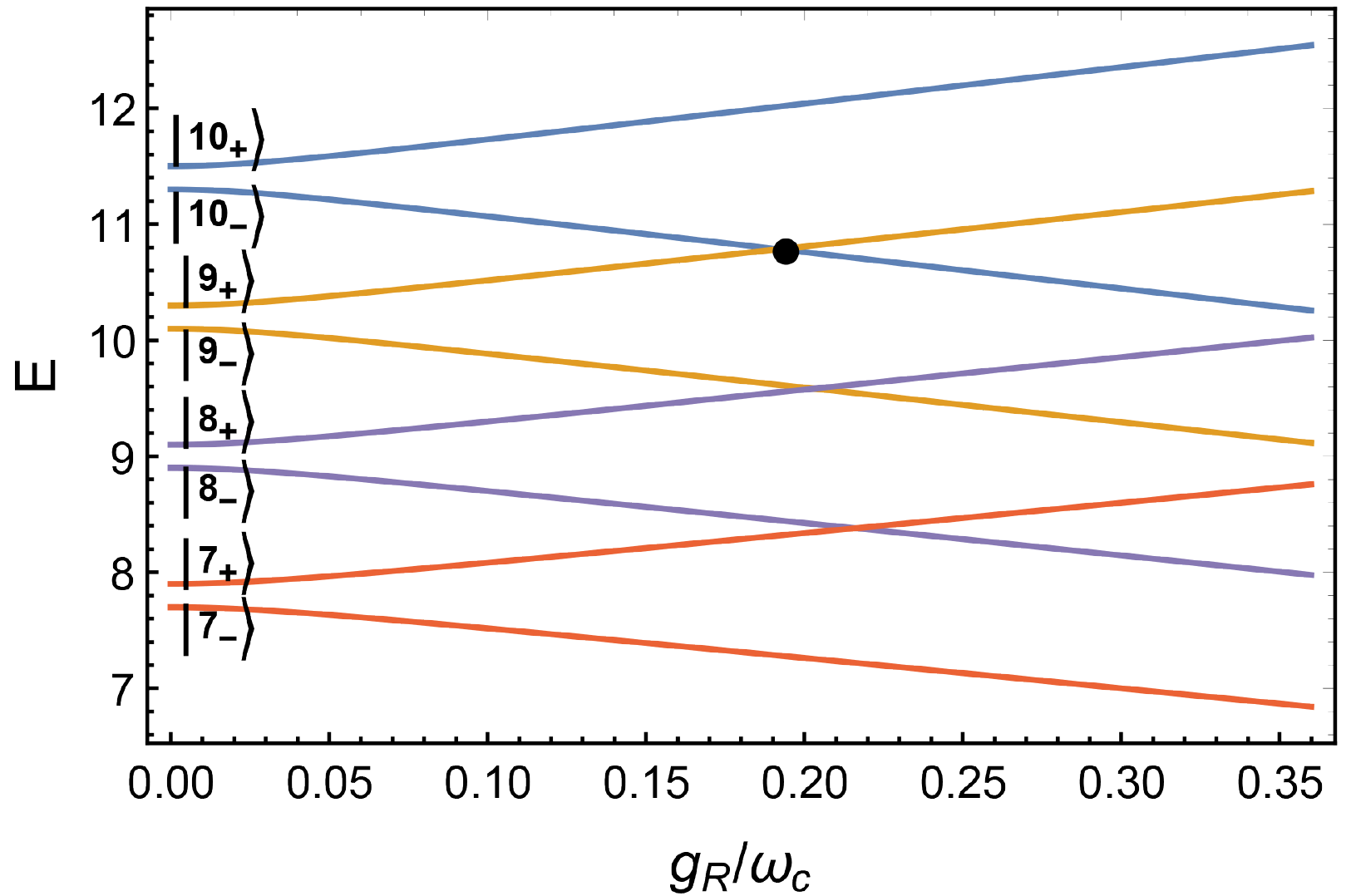}}
\caption{Energy spectrum in Eq.~\eqref{energieslevel} versus $g_{R}/\omega_{c}$
with $n=7\div10$, for (a) the resonant case $\omega_{a}=\omega_{c}$, and (b)
far from resonance $\left(\omega_{c}-\omega_{a}\right)/\omega_{c}=0.2$. The black dots around $g_{R}=0.16\,\omega_{c}$
in (a) or $g_{R}=0.19\,\omega_{c}$ in (b) indicate the first energy
crossings according to the JC model. For coupling constants beyond
these values the structure of the energy ladder from Fig.~\ref{fig:transitions}
is not preserved. }
\label{fig:energy_10_resonance}
\end{figure*}
The transition frequencies will naturally depend on the coupling strength
$g_{R}$, as in the Jaynes-Cummings theory, and
weakly on $g_{S}$ through second order perturbation [Eq.~(\ref{eq:energy_2nd_order})].
The Jaynes-Cummings energy structure is shown in Fig. \ref{fig:energy_10_resonance}
for manifolds $n=7$ to $n=10$, both in the resonant
$\omega_{c}=\omega_{a}$ and detuned case $\omega_{c}-\omega_{a}=0.2\,\omega_{c}$.

\begin{figure*}
\includegraphics[width=0.48\textwidth]{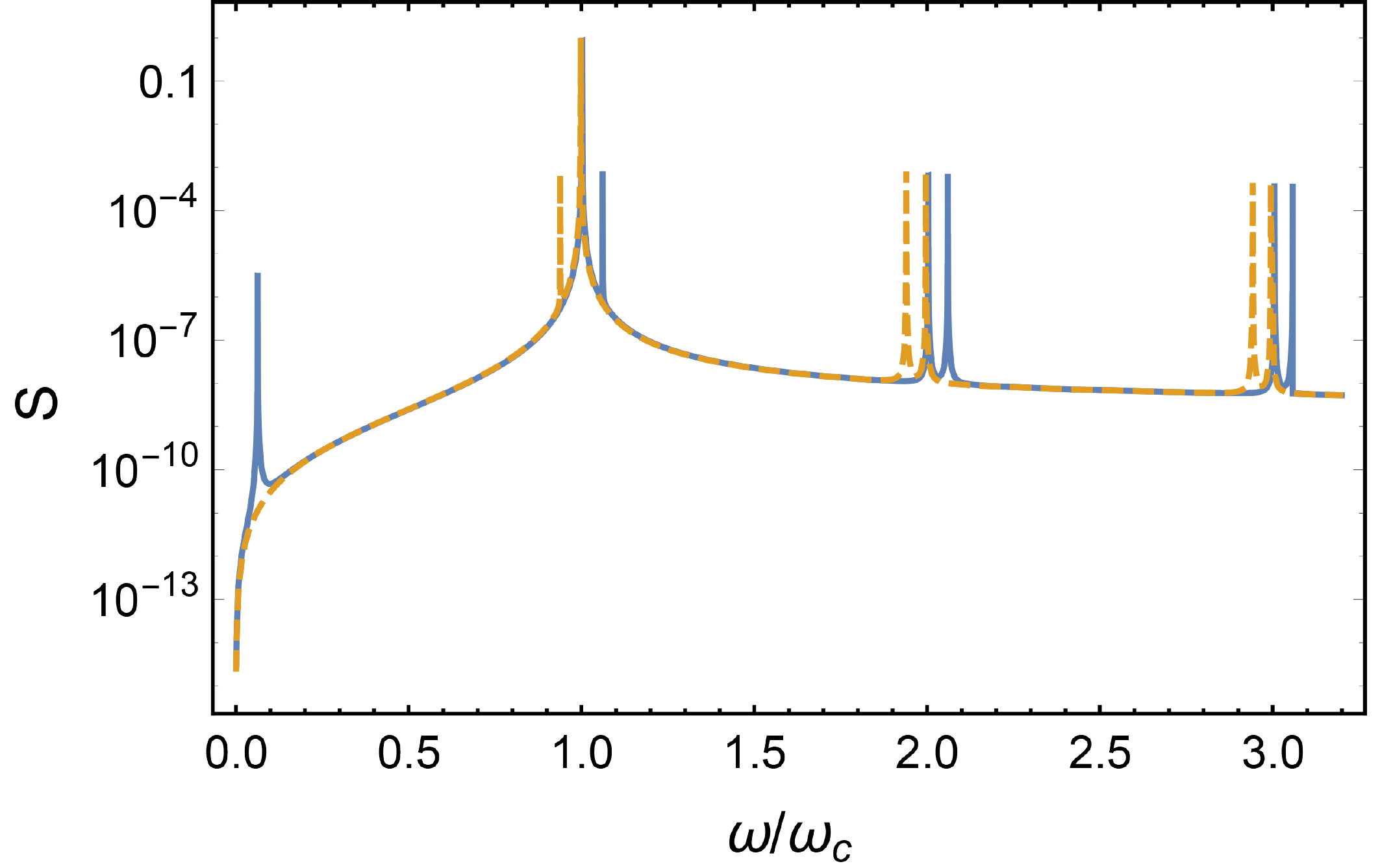}\qquad\includegraphics[width=0.48\textwidth]{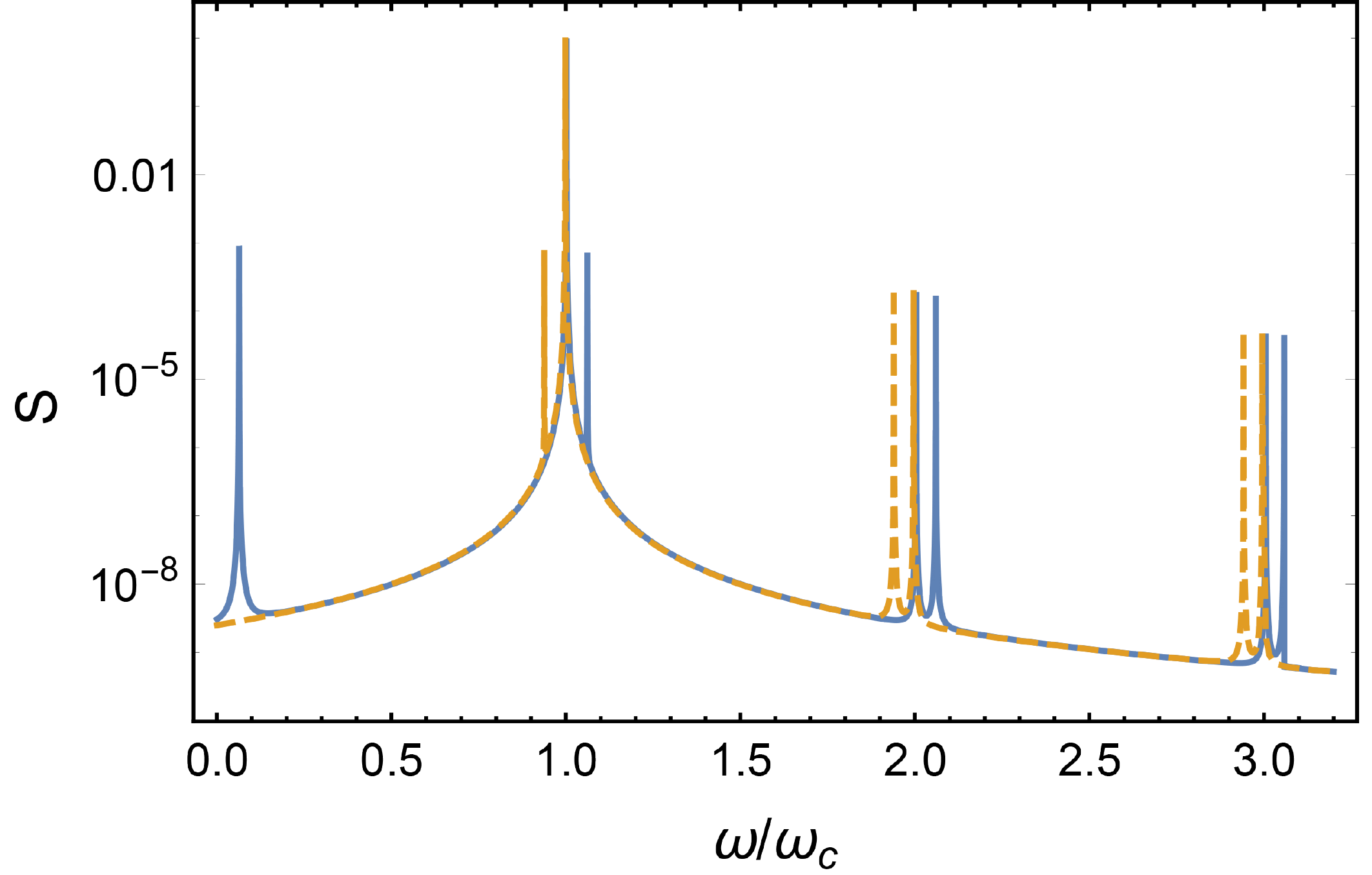}
\caption{Emission spectra (arbitrary units) from the initial states $\left|10_{+}\right\rangle$ (blue) and $\left|10_{-}\right\rangle $ (orange), with form factors scaling like $\omega^{p}$, with (a) $p=2$, and (b) and $p=0$. Plots are normalized to the maximum of $S$. }\label{fig:spec} 
\end{figure*}

According to Eq.~\eqref{distribution}, the emission spectrum is approximately made up of a set of Lorentzian peaks. In Fig.~\ref{fig:spec} we
separately plot the spectra for the initial states $\left|10_{+}\right\rangle $
(solid blue line) and $\left|10_{-}\right\rangle $ (dashed orange
line), for the resonant case $\omega_{a}=\omega_{c}$ and fixed coupling
strengths $g_{R}=g_{S}=0.01\,\omega_{c}$. The spectra are plotted for
two different form factors, $\mathcal{P}\propto\omega^{2}$ [Fig.~\ref{fig:spec}(a)] and $\mathcal{P=\mathrm{const.}}$ [Fig.~\ref{fig:spec}(b)], corresponding respectively to three- and one-dimensional reservoir geometries in the case of frequency-independent coupling between cavity and environment.
The single low-energy peak around $\omega=2\sqrt{10}g_{R} = 0.063\,\omega_{c}$
corresponds to the $\left|10_{+}\right\rangle \rightarrow\left|10_{-}\right\rangle $
transition induced by the inversion-symmetry breaking of the two-level
system and might unveil applications for low-frequency-sources. Therefore,
its tunability is an important feature: the position of this peak
depends on $g_{S}$, i.e.\ on the permanent dipole moment
$\boldsymbol{d}_{\mathrm{ee}}$ of the atom, and on the field strength in the cavity, related to the number of photons. In the classical limit, this provides
an all-optical tuning possibility with the field amplitude~\cite{kibis2009}. Additionally,
tuning could be achieved through orientation of the permanent dipole
moment of the two-level system with an external DC electric field
\cite{gladysz2020}. Around $\omega=\omega_{c}$ we recognize the Mollow
triplet that arises from the JC interaction. Similar structures are
repeated around $\omega=2\,\omega_{c}$ and $\omega=3\,\omega_{c}$, arising
respectively from the AS and CR Hamiltonian perturbations. Note that
the positions of sidebands of the Mollow-like triplet around $2\,\omega_{c}$
are related to the diagonal dipole moment and will accordingly be
modified if $g_{S}$ is tuned. We emphasize that all the peaks, including
the Mollow-like sidebands, can be resolved in the spectra. In particular,
even though the low-energy peak usually corresponds to the weakest
transition intensities, it appears on top of a correspondingly suppressed
background. As a consequence, the signal-to-noise ratio is found comparable
for all emission peaks. Below we analyze the intensity ratio of different
peaks depending on the coupling strengths of the model. 

\begin{figure*}[t!]
\subfigure[$\mathcal{P}=\mathrm{const.}$, resonant]{\includegraphics[width=0.32\textwidth]{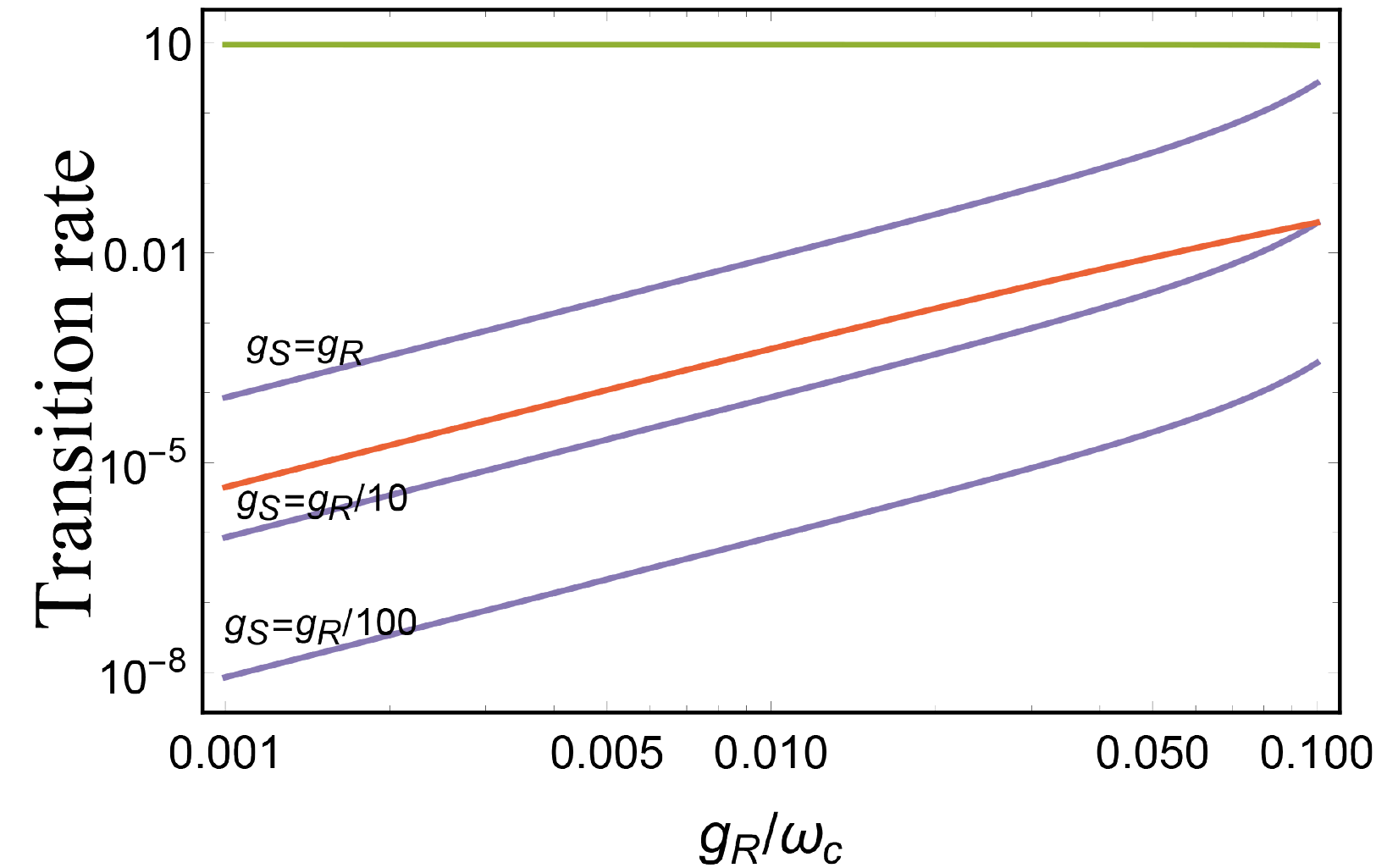}
}\hfill
\subfigure[$\mathcal{P}\propto \omega^2$, resonant]{\includegraphics[width=0.32\textwidth]{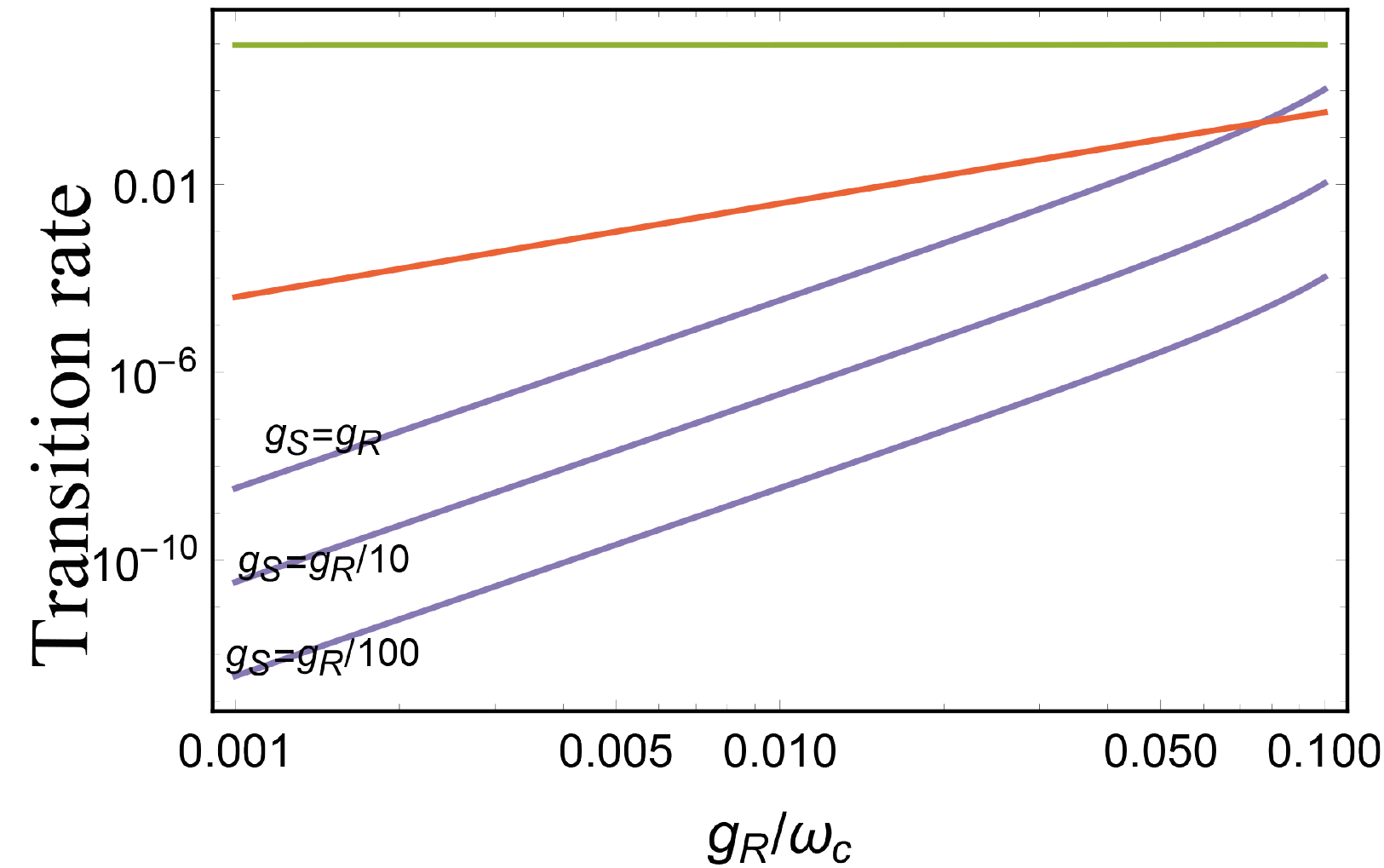}
}\hfill
\subfigure[Lorentzian $\mathcal{P}$, resonant]{\includegraphics[width=0.32\textwidth]{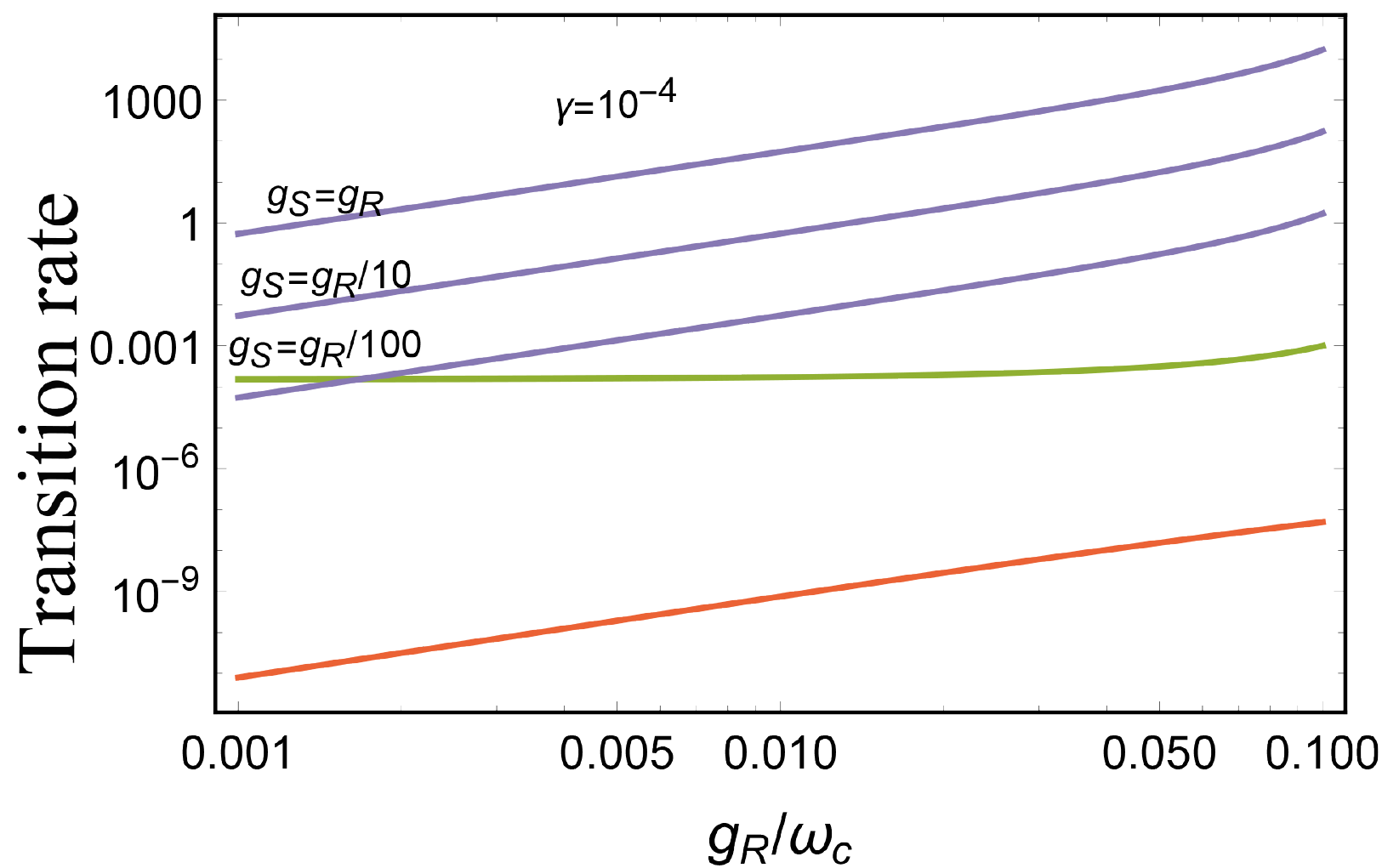}
}
\newline
\subfigure[$\mathcal{P}=\mathrm{const.}$, detuned]{\includegraphics[width=0.32\textwidth]{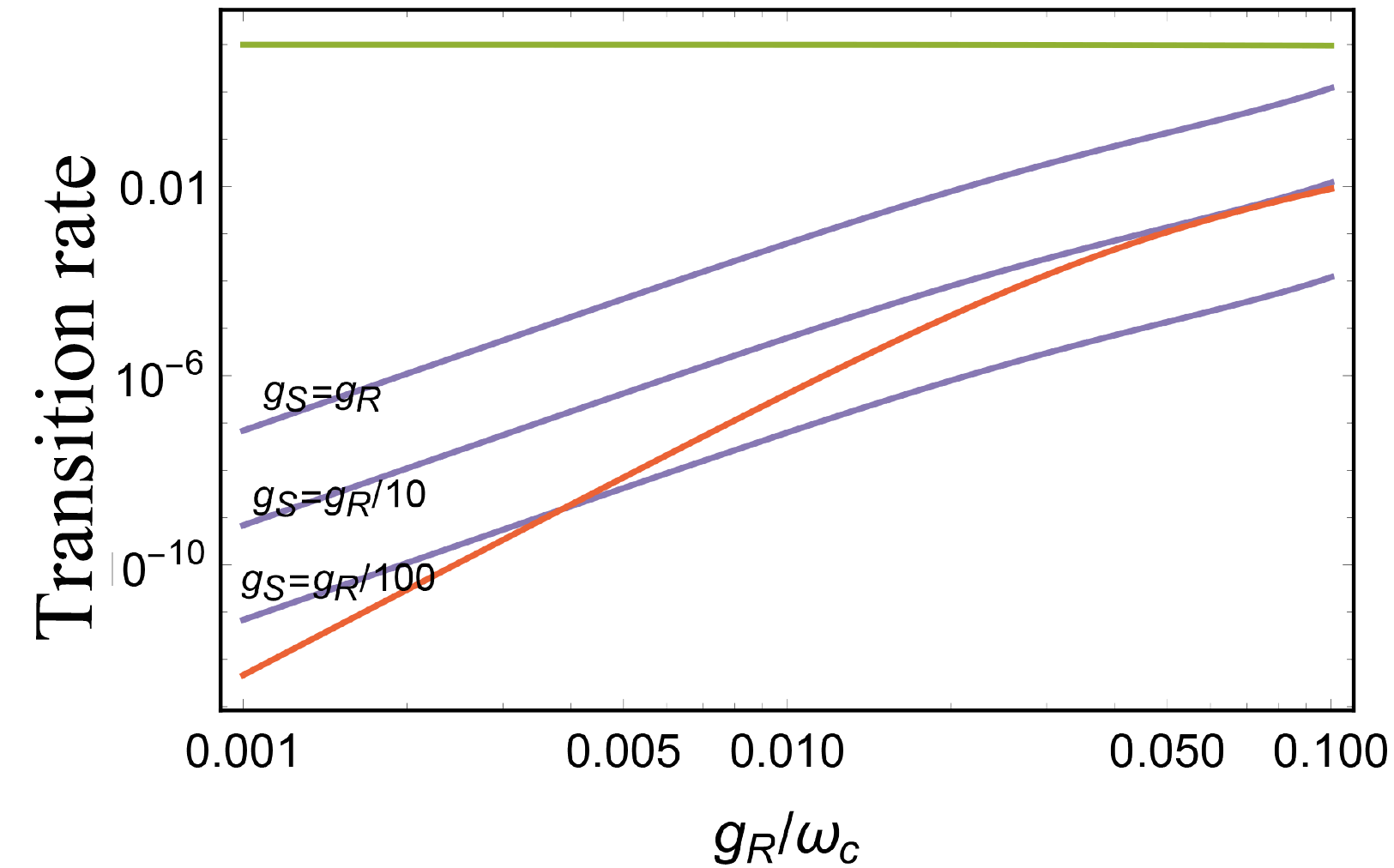}
}\hfill
\subfigure[$\mathcal{P}\propto \omega^2$, detuned]{\includegraphics[width=0.32\textwidth]{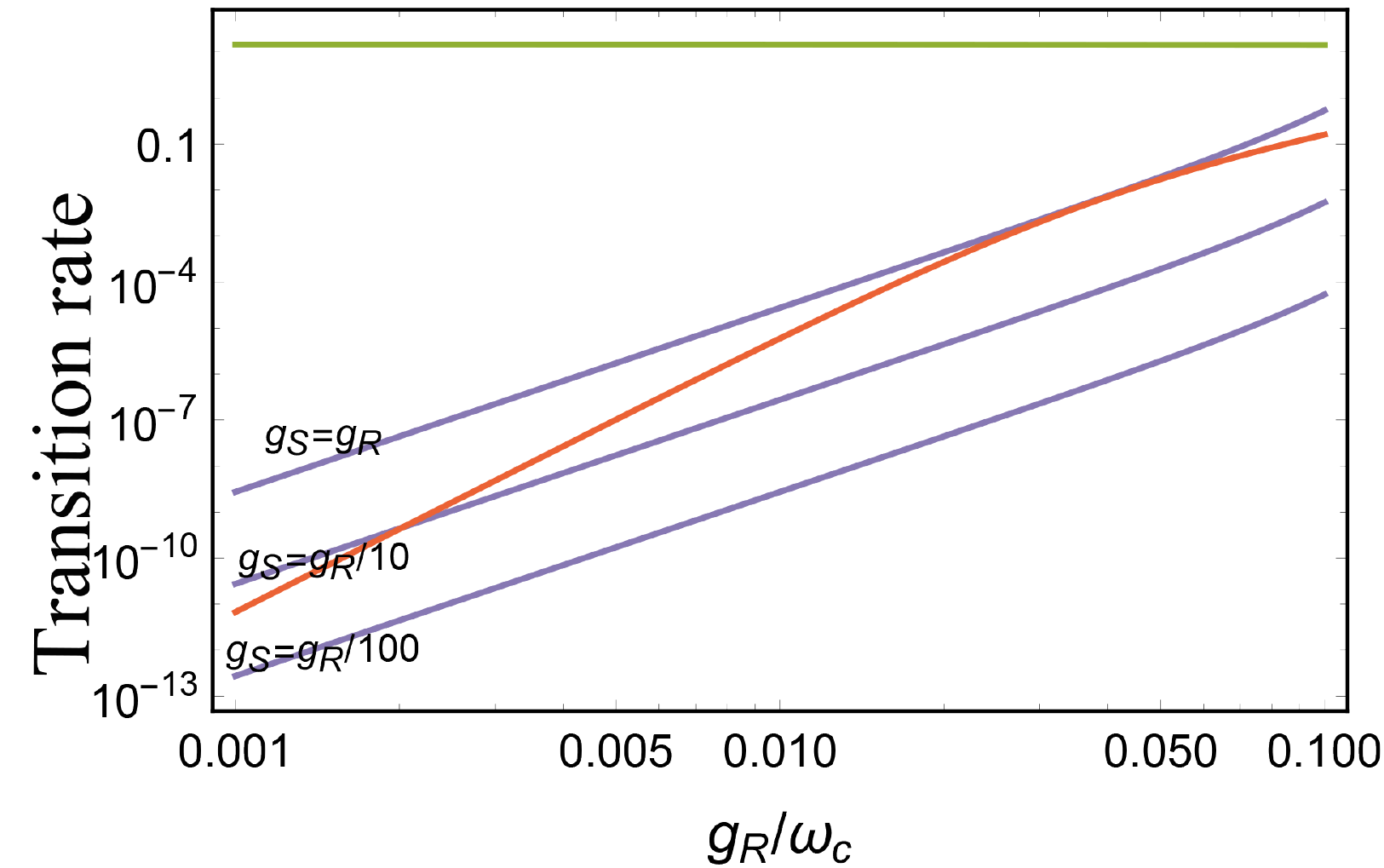}
}\hfill
\subfigure[Lorentzian $\mathcal{P}$, detuned]{\includegraphics[width=0.32\textwidth]{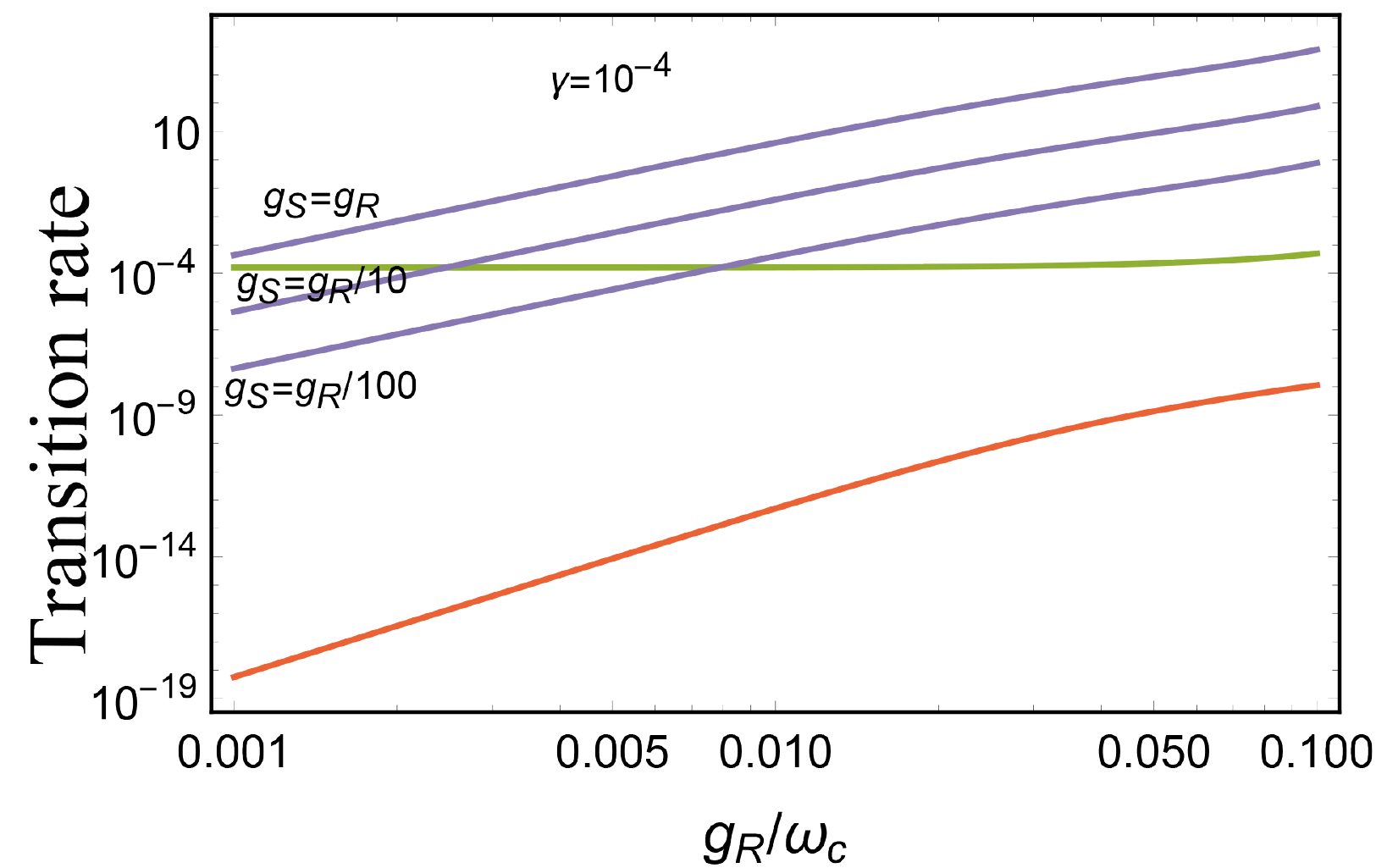}
}
\caption{
Transition rates determined by a form factor $\mathcal{P}=\mathrm{const.}$ [panels
(a) and (d)], $\mathcal{P}\propto\omega^2$ [panels (b) and (e)],
and Lorentzian, centered at the frequency $\omega_{\mathrm{ext}}=E_{nn}^{+-}/\hbar$,
with a quality factor $\gamma_{\mathrm{ext}}=10^{-4}\omega_{\mathrm{ext}}$  [panels (c) and (f)]. Results are referred to the resonant $\omega_{c}=\omega_{a}$ [panels (a), (b) and (c)] and detuned case $\left(\omega_{c}-\omega_{a}\right)/\omega_{c}=0.2$ [panels (d), (e) and (f)]. Plots are in arbitrary units, as only ratios between different rates are relevant for our analysis.
In panels (a) and (d), the plotted quantities correspond, up to a constant, to the squared matrix elements $\left|\left\langle {n'}_{s'}|a|n_s\right\rangle \right|^{2}$.
Colors indicate different transition mechanisms: the Jaynes-Cummings
transition $\left|10_{+}\right\rangle \to\left|9_{+}\right\rangle $
is shown in green, the counter-rotating term $\left|10_{+}\right\rangle \to\left|7_{-}\right\rangle $
in red, and the transition driven by the diagonal coupling $\left|10_{+}\right\rangle \to\left|10_{-}\right\rangle $
in violet, for 3 values of the diagonal coupling strength, $g_{S}=g_{R},g_{R}/10,g_{R}/100$.
 \label{fig:amplitudes}}
\end{figure*}
We study three selected transitions, representative for each Hamiltonian
contribution, highlighted as thick arrows in Fig.~\ref{fig:transitions}:
for the Jaynes-Cummings term we select the $|10_{+}\rangle\rightarrow|9_{+}\rangle$
transition; for the diagonal coupling term  the $|10_{+}\rangle\rightarrow|10_{-}\rangle$
transition; for the counter-rotating term the $|10_{+}\rangle\rightarrow|7_{-}\rangle$
transition, that corresponds to the highest frequency. In Fig. \ref{fig:amplitudes}(a),
we plot the squared matrix elements $\left|\left\langle {n'}_{s'}|a|n_s\right\rangle \right|^{2}$, which entirely determine the relative weight of the different decay channels in the case of a constant form factor [see Eq.~(\ref{weights})]. They are
plotted separately for each considered transition. As anticipated,
the contribution due to the Jaynes-Cummings interaction dominates, overcoming the other terms by several orders of magnitude for the
investigated range of coupling strengths $g_{R}$. As expected, the JC contribution has a relatively weak dependence on $g_{R}$, which induces small corrections to the zeroth-order result. The purple (red) lines in Fig.~\ref{fig:amplitudes}
represent the contributions determined by the AS (CR) Hamiltonian. Results obtained for the different
values $g_{S}=g_{R},\:g_{R}/10,\:g_{R}/100$ are
presented. This confirms the intuition suggested at the end
of the previous section, that for equal coupling strengths $g_{S}=g_{R}$ the
term induced by the diagonal coupling overcomes the counter-rotating
contribution. Both terms share the same linear scaling with their
respective coupling strengths $g_{S}$ or $g_{R}$, so, as
we decrease $g_{S}$, the squared transition amplitude $\left|\left\langle {n}_{s'}|a|n_s\right\rangle \right|^{2}$
is gradually suppressed. 

This simple linear scaling is slightly modified in the detuned case,
in which the slopes change around \mbox{$g_{R}\simeq (\omega_{c}-\omega_{a})/2\sqrt{n}$}.
An example for a strong detuning \mbox{$\omega_{a}=0.8\,\omega_{c}$} is shown
in Fig.~\ref{fig:amplitudes}(d). We find that in this case the contribution
of both perturbative terms is suppressed with respect to the resonant
contribution. However, for relatively small coupling strengths ($g_{R}<4\times10^{-3}\,\omega_{c}$)
the terms corresponding to the asymmetric contribution still dominate
over those due to the counter-rotating Hamiltonian, even for small
$g_{S}=0.01g_{R}$.

For a wide range of coupling strengths, the squared transition
amplitudes induced by the perturbation related to the
asymmetry dominate over those originating from the counter-rotating
term. However, if the outcoupling Hamiltonian $H_{\mathrm{ext}}$ involves a form factor scaling as $\omega^p$, the weight of a decay channel is proportional to the $p$-th power of the transition frequency.
Therefore, in a 3D continuum geometry, in which the density of states scales as $\omega^2$, the relevance of low-energy transitions tends to be suppressed. We show this case in
both the resonant and off-resonant case in Fig.~\ref{fig:amplitudes}(b) and (e).
In the off-resonant case, we note that for equal coupling strengths
$g_{S}=g_{R}$ the terms originating from the diagonal-coupling still
dominate over the counter-rotating ones, despite the latter being by
far energetically favored. 

The different behavior of transition rates in the cases of constant $\mathcal{P}$
{[}Fig.~\ref{fig:amplitudes}(a) and (d){]} and $\mathcal{P}\propto\omega^2$
{[}Fig.~\ref{fig:amplitudes}(b) and (e){]} suggests the possibility
of tailoring the output by engineering the coupling to the continuum and its density of states. To further highlight this point, we couple the atom-cavity system to a single-mode cavity, assuming that the form factor $\mathcal{P}$ is a Lorentzian function, centered at the
low-energy transition frequency \mbox{$\hbar\omega_{\mathrm{ext}}=E_{nn}^{+-}$}
and characterized by a full-width at half-maximum $\gamma_{\mathrm{ext}}=10^{-4}\omega_{c}.$
A cavity with similar parameters can be realized in photonic crystals~\cite{lu2018}
that provide 1D or 2D photonic environments, in whispering-gallery-mode
resonators~\cite{vogt2018} or, with smaller quality factors,
using meta-materials~\cite{zografopoulos2015}. Here, the cavity
is tailored to emphasize the strength of the low-energy transition
$|n_{+}\rangle\rightarrow|n_{-}\rangle$ at the cost of suppressing
other transitions. Indeed, as demonstrated in Fig.~\ref{fig:amplitudes}(c) and (f),
this is successful in both the resonant and detuned case. 

\begin{figure*}[t!]
\subfigure[1D, resonant]{\includegraphics[width=0.49\textwidth]{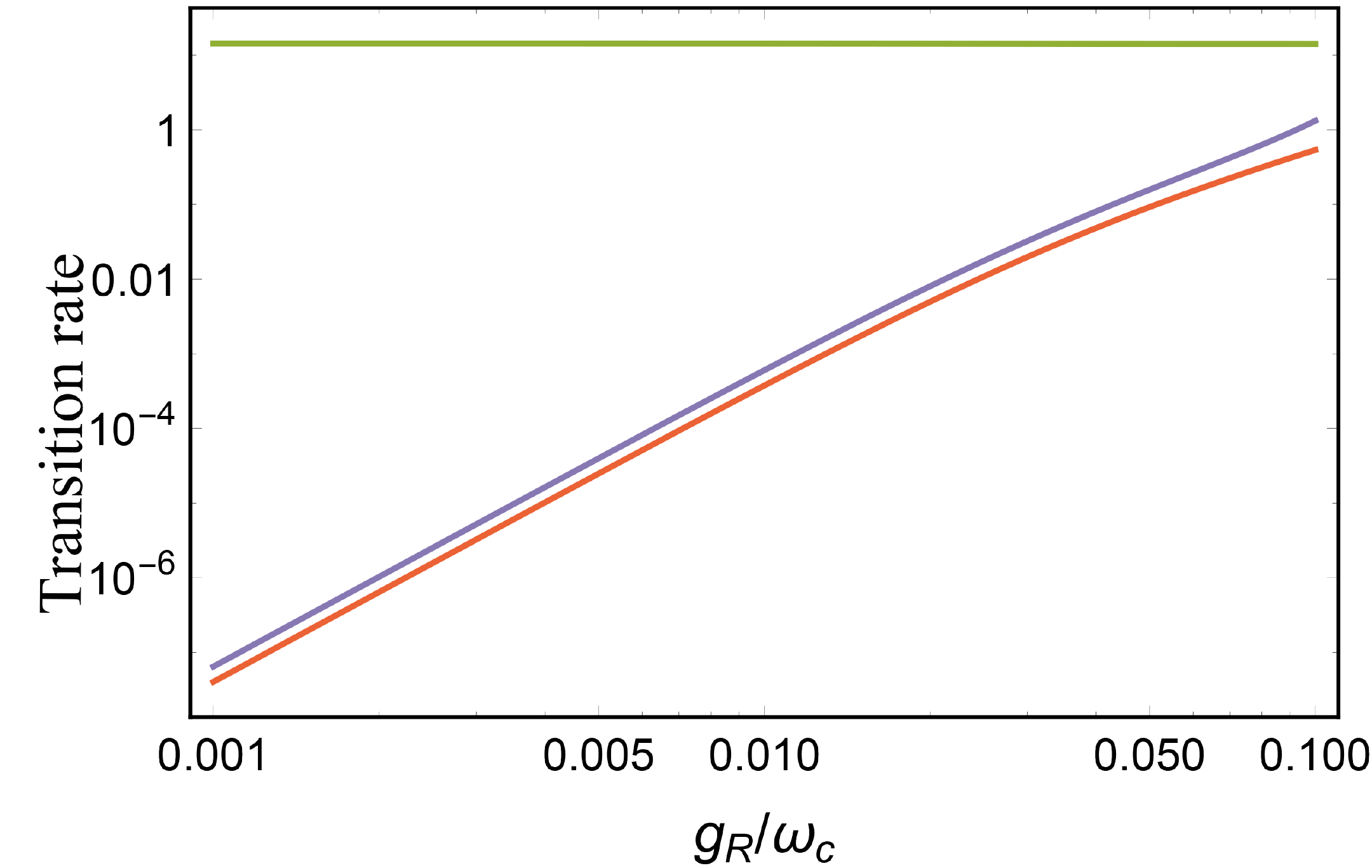}}\hfill
\subfigure[1D, detuned]{\includegraphics[width=0.49\textwidth]{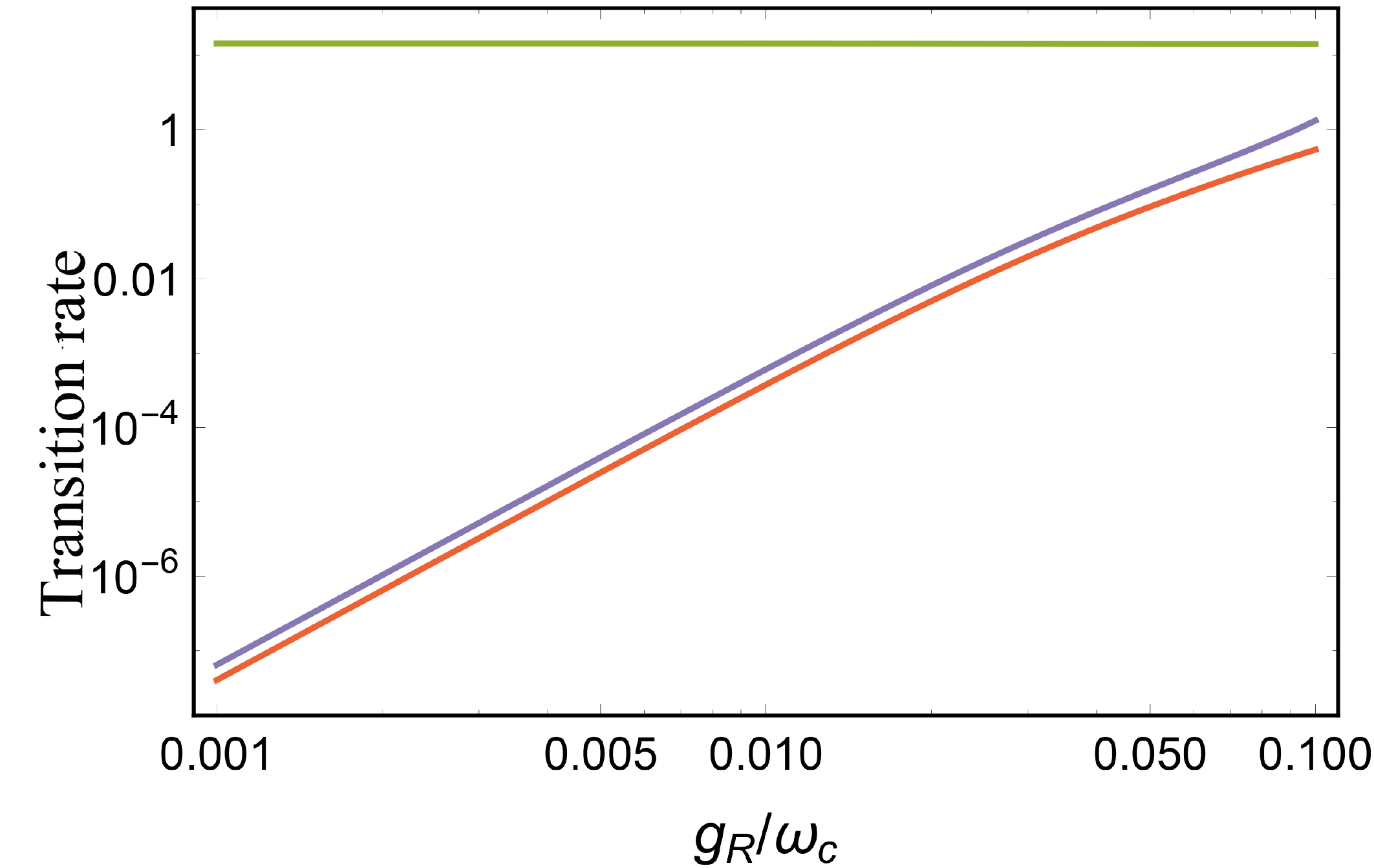}}
\newline
\subfigure[resonant]{\includegraphics[width=0.49\textwidth]{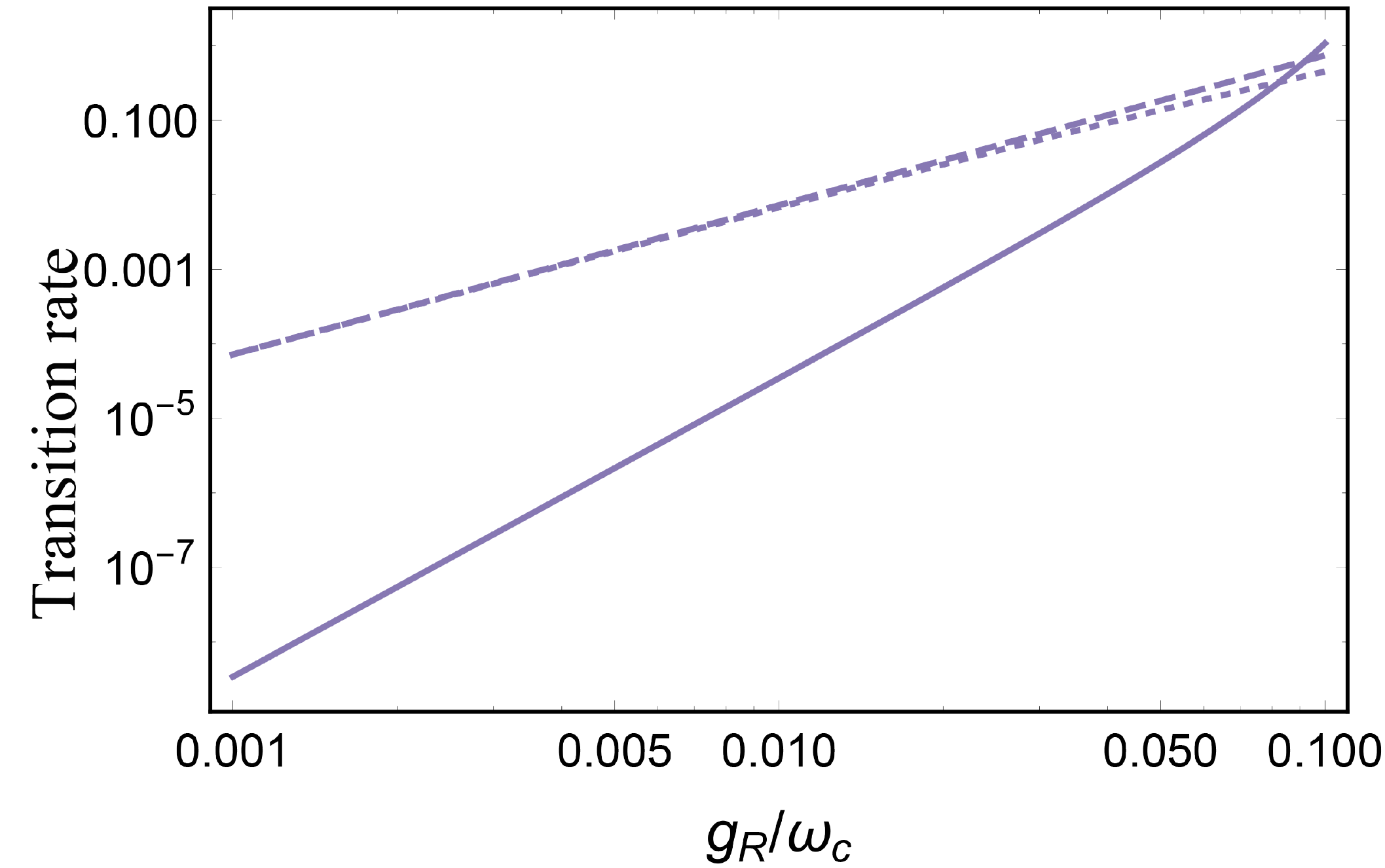}}\hfill
\subfigure[detuned]{\includegraphics[width=0.49\textwidth]{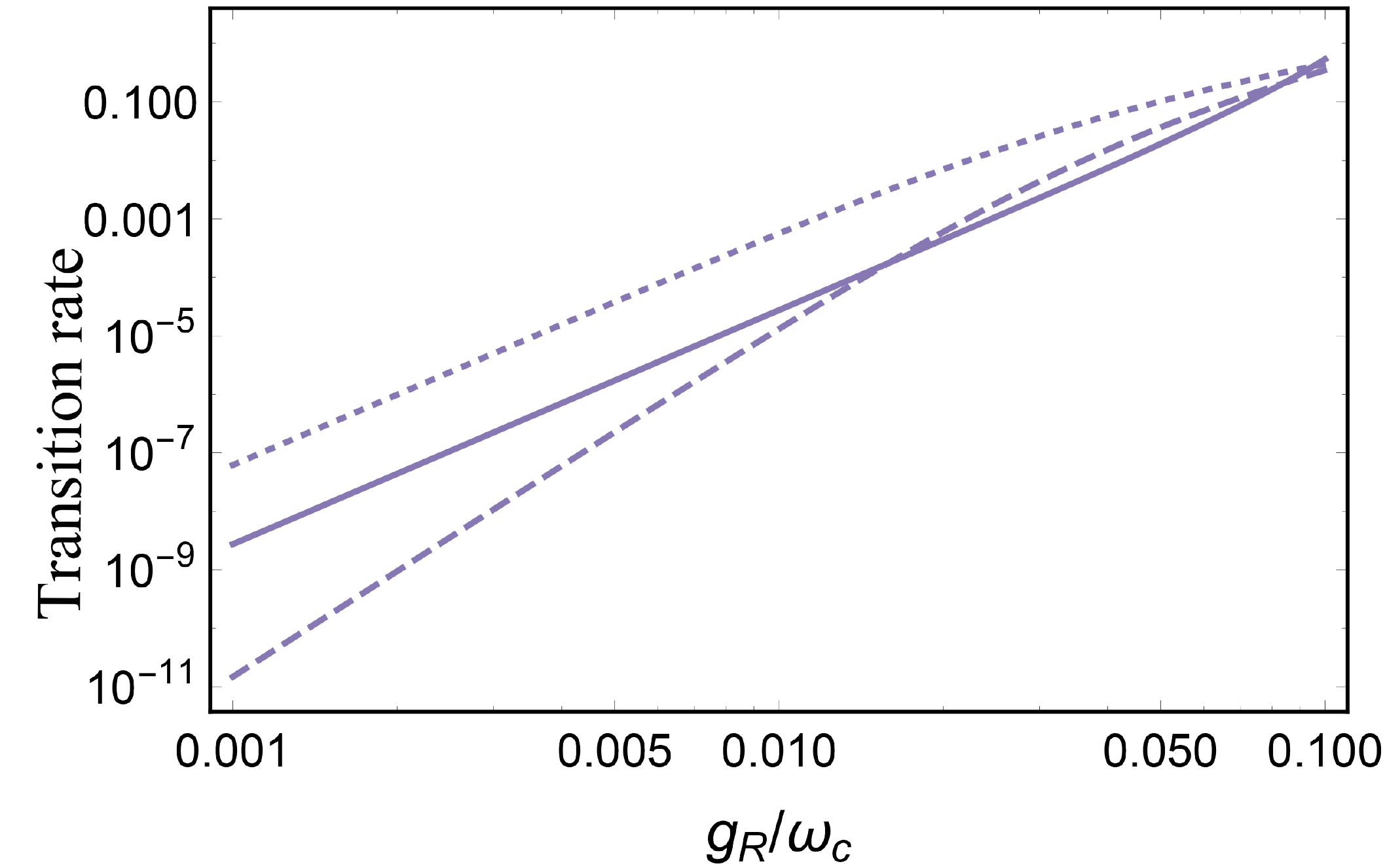}}
\caption{
Total transition rates for each Hamiltonian contribution, for a fixed initial state $\left|10_{+}\right\rangle $: rate of the Jaynes-Cummings
transitions $\Gamma^{\mathrm{JC}}_{10,+}=\Gamma_{10,9}^{++}+\Gamma_{10,9}^{+-}$ (green), diagonal coupling
mechanism $\Gamma^{\mathrm{AS}}_{10,+}=\Gamma_{10,8}^{++}+\Gamma_{10,8}^{+-}+\Gamma_{10,10}^{+-}$ (violet),
and the counter-rotating transitions $\Gamma^{\mathrm{CR}}_{10,+}=\Gamma_{10,7}^{++}+\Gamma_{10,7}^{+-}$
(red) (a) on resonance, and (b) for the detuned case.
Individual contributions 
to the diagonal coupling are resolved in panels (c) and (d) for the resonant
and the detuned case, respectively, where the solid line represents the low-energy
transition rate $\Gamma_{10,10}^{+-}$, while the dashed ($\Gamma_{10,8}^{++}$)
and dotted ($\Gamma_{10,8}^{+-}$) lines correspond to transitions
around $2\,\omega_{c}$. 
\label{fig:grouped}}
\end{figure*}
For the above analysis we have selected only one exemplary transition
of the Jaynes-Cummings, diagonal-coupling and counter-rotating groups,
corresponding to  arrows with  different colors in Fig.~\ref{fig:transitions}.
In Fig.~\ref{fig:grouped}, we show the total transition rates in each group, considering the initial state $|n_s\rangle=|10_{+}\rangle$. The green lines correspond to the
total rate of the Jaynes-Cummings transitions $\Gamma^{\mathrm{JC}}_{10,+}=\Gamma_{10,9}^{++}+\Gamma_{10,9}^{+-}$, the purple
lines to the diagonal coupling $\Gamma^{\mathrm{AS}}_{10,+}=\Gamma_{10,8}^{++}+\Gamma_{10,8}^{+-}+\Gamma_{10,10}^{+-}$,
and the red lines to the counter-rotating contribution $\Gamma^{\mathrm{CR}}_{10,+}=\Gamma_{10,7}^{++}+\Gamma_{10,7}^{+-}$. 
We find that, as expected, the higher-energy contributions from the
asymmetric Hamiltonian $H_{\mathrm{AS}}$ around $2\,\omega_{c}$ are strong enough to overcome the ones induced by the counter-rotating terms. This can be also seen from Fig.~\ref{fig:grouped}(c), in which we resolve different contributions induced by $H_{\mathrm{AS}}$
in the decay from the state $|10_{+}\rangle$. The
difference between the two perturbative contributions becomes even
smaller in the detuned case, in which all perturbative terms are
suppressed, as can be seen from panels (b) and (d) in Fig.~\ref{fig:grouped}.

\section{Conclusions}

We have applied second-order perturbation theory to investigate the
emission properties of a two-level system coupled to a single-mode
electromagnetic field, including interaction channels based on the
Jaynes-Cummings, counter-rotating and asymmetry-related contributions.
In the electric-dipole interaction mechanism, the first two interactions arise
from the coupling of the field mode with the induced transition dipole
moment, while the latter requires a permanent dipole characterizing the
system's eigenstates. Light-matter coupling with permanent dipoles
gives birth to additional emission peaks. We have demonstrated that
even though at some frequencies the asymmetry-related contribution
is weak in relative terms, the signal-to-noise ratio is comparable
for all emission peaks. Moreover, the relative strengths of the emission
peaks can be modified with a suitable photonic environment, as we
have discussed for 1D systems and for a Lorentzian cavity. In the
latter example we have shown that for cavity parameters that lie well
within the range of experimental capabilities, the asymmetry-related
emission channel may even become dominant. 
\begin{acknowledgments}
We acknowledge the PROM project at the Nicolaus Copernicus University in Toru\'{n}. 
GS is supported by The International Centre for Theory of Quantum Technologies project (contract no. 2018/MAB/5) carried out within the International Research Agendas Programme of the Foundation for Polish Science co-financed by the European Union from the funds of the Smart Growth Operational Programme, axis IV: Increasing the research potential (Measure 4.3). KS is supported by the Polish National Science Centre project
2018/31/D/ST3/01487. 
PF and SP acknowledge support by MIUR via PRIN 2017 (Progetto di Ricerca di Interesse Nazionale), project QUSHIP (2017SRNBRK). PF was partially supported by the Italian National Group of Mathematical Physics (GNFM-INdAM). 
 PF, SP, and FVP were partially supported by Istituto Nazionale di Fisica Nucleare (INFN) through the project ``QUANTUM'' and by Regione Puglia and  QuantERA ERA-NET Cofund in Quantum Technologies (GA No.\ 731473), project PACE-IN. 
\end{acknowledgments}

\appendix
\section{Validity of the perturbative approach}\label{app:test}

In this Appendix we discuss the range of coupling strengths for which
the perturbative approach used in the main text is justified. The condition for the analysis to be consistent is that the perturbation
series converge both for the perturbed energies and states. This
is not the case around energy crossings, where some of the series
terms in the perturbed eigenstate given by Eq.~(\ref{eq:ket_2nd_order}) diverge. On the other hand, sufficiently away from energy crossing the state norm is approximately preserved. 
To identify the applicability range of the approach,
we therefore verify the normalization of states. 

\begin{figure}[h]
\includegraphics[width=0.48\textwidth]{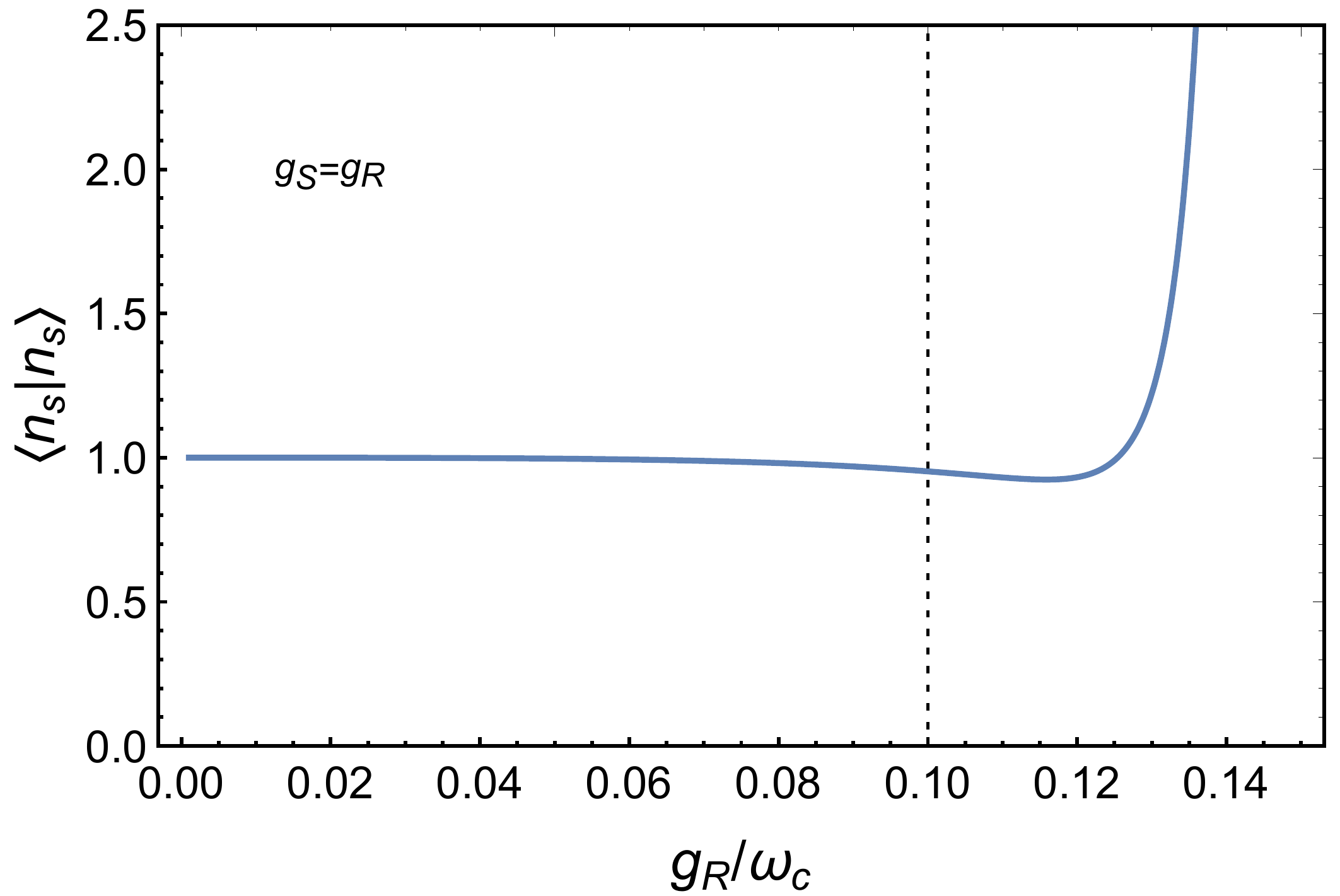}
\caption{Norm of state $|10_{+}\rangle$, as obtained from Eq.~\eqref{eq:ket_2nd_order}. For coupling strengths exceeding $g_{R} \simeq 0.1\,\omega_{c}$ the norm deviates from 1, which is
an indication of the breakdown of perturbation theory. }\label{fig:norm}
\end{figure}

For the cases investigated
in the main manuscript, the first energy crossing among the investigated
states appears for those corresponding to the highest manifold. In
Fig. \ref{fig:norm}, we plot the norm of state $\left|10_{+}\right\rangle $
as a function of $g_{S}=g_{R}$. A clear divergence appears for coupling strengths approaching $0.14\,\omega_{c}$, which results from
a crossing involving higher manifolds, in this case up to $\mathcal{E}\left(14\right)$.
The vertical line in the figure indicates the limit for the coupling
strengths considered in the main text. 

\section{Spectral distribution of emitted photons}\label{app:resolvent}

In order to derive Eqs.~\eqref{Gamma}-\eqref{distribution}, let us consider an initial state $\ket{\psi_0}$, evolving under the Hamiltonian $H=H_0+H_{\mathrm{int}}$, where $H_{\mathrm{int}}$ is meant as a perturbation of the ``free'' Hamiltonian $H_0$. While 
\begin{equation}
H_0 \ket{\psi_0} = E_0 \ket{\psi_0},
\end{equation}
the presence of the interaction Hamiltonian $H_{\mathrm{int}}$ makes the initial state unstable, inducing decay towards generalized eigenstates $\ket{q}$ of $H_0$ (where $q$ is generally a multi-index of $N$ quantum numbers, some of which can be discrete, such as spin or polarization), satisfying
\begin{equation}
H_0 \ket{q} = E(q) \ket{q} .
\end{equation}
The probability associated to a specific $q$ at an arbitrary time $t$ can be computed by projecting $\ket{q}$ on the evolved state $\ket{\psi(t)}=\exp(-\ii H t)\ket{\psi}$ of the system, determined by the corresponding matrix element of the resolvent \mbox{$(z-H)^{-1}$} through a Fourier-Laplace transform:
\begin{equation}\label{fourierlaplace}
\mathcal{A}(q,t) = \bra{q} \ee^{-\ii H t/\hbar}\ket{\psi_0} = \frac{\ii}{2\pi} \int_{\mathcal{B}} dz\, \ee^{-\ii z t/\hbar} \bra{q} \frac{1}{z-H} \ket{\psi_0} ,
\end{equation}
where $\mathcal{B}=(-\infty+\ii \eta,+\infty+\ii \eta)$ is an arbitrary line, parallel to the real axis, with $\eta>0$. Therefore, in order to characterize the matrix element of the resolvent, it is sufficient to determine the amplitude and probability associated to the distribution of decay products. In particular, we are interested in the asymptotic distribution
\begin{equation}
P_{\infty}(q) = \lim_{t\to\infty} \left| \bra{q} \ee^{-\ii H t/\hbar}\ket{\psi_0} \right|^2 .
\end{equation}
For $\mathrm{Im}(z)\neq 0$, the resolvent satisfies the equation
\begin{equation}
\frac{1}{z-H} = \frac{1}{z-H_0} + \frac{1}{z-H_0} H_{\mathrm{int}} \frac{1}{z-H} ,
\end{equation}
that can be used, along with the assumption $\bra{\psi_0}H_{\mathrm{int}}\ket{\psi_0}$, to determine an approximate form of the matrix element appearing in the right-hand side of~\eqref{fourierlaplace},
\begin{equation}\label{lie}
\bra{q} \frac{1}{z-H} \ket{\psi_0} \simeq \frac{1}{z-E(q)} \bra{q}H_{\mathrm{int}}\ket{\psi_0} \bra{\psi_0} \frac{1}{z-H} \ket{\psi_0}  ,
\end{equation}
where the corrections are $O(H_{\mathrm{int}}^2)$ and proportional to the matrix elements $\bra{q}H_{\mathrm{int}}\ket{q'}$. An error of the same order on $\mathcal{A}(q,t)$ is entailed by applying the Weisskopf-Wigner approximation to the initial state propagator,
\begin{equation}
\bra{\psi_0} \frac{1}{z-H} \ket{\psi_0} \simeq \frac{1}{z-(E_0 + \hbar\Delta - \ii \hbar\Gamma/2)}, 
\end{equation}
with
\begin{align}
\Delta & = \frac{1}{\hbar} \mathrm{P} \int d^N q\, \frac{\left| \bra{q}H_{\mathrm{int}}\ket{\psi_0} \right|^2}{E_0 - E(q)} , \\
\Gamma & = \frac{2\pi}{\hbar} \int d^N q\, \left| \bra{q}H_{\mathrm{int}}\ket{\psi_0} \right|^2 \delta(E_0 - E(q)) ,
\end{align}
which yields an expression of the transition amplitude in terms of a solvable integral
\begin{equation}
\mathcal{A}(q,t) \simeq \frac{\ii}{2\pi} \int_{\mathcal{B}} dz\,  \frac{\bra{q}H_{\mathrm{int}}\ket{\psi_0} \ee^{-\ii z t/\hbar} }{(z-E(q))(z-(E_0 + \hbar\Delta - \ii \hbar\Gamma/2))} ,
\end{equation}
leading to the asymptotic distribution
\begin{equation}\label{probWW}
P_{\infty}(q) \simeq \frac{ \left| \bra{q}H_{\mathrm{int}}\ket{0} \right|^2 }{(E(q) - E_0 - \hbar\Delta)^2 + \hbar^2\frac{\Gamma^2}{4}} .
\end{equation}
Suppose now that the final states $\ket{q}$ can be collected in different decay channels, namely orthogonal subspaces $\mathcal{D}_j$ of final products, identified by quantum numbers belonging to specific domains $D_j$. The energy distribution associated to the decay channel $n$ reads 
\begin{align}\label{prob}
P^{(j)}_{\infty}(E) & = \int_{D_j} d^Nq\, \delta(E-E(q)) P_{\infty}(q) \nonumber \\ & \simeq \frac{1}{2\pi} \frac{ \hbar\Gamma_j }{(E - E_0 - \hbar\Delta)^2 + \hbar^2\frac{\Gamma^2}{4}} ,
\end{align}
with 
\begin{equation}
\Gamma_j = \frac{2\pi}{\hbar} \int_{D_j} d^Nq\, \left| \bra{q}H_{\mathrm{int}}\ket{\psi_0} \right|^2 \delta(E_0 - E(q)) ,
\end{equation}
the channel decay rate. Computation of the total probability for the system to decay in channel $n$ yields the classical result
\begin{equation}
p^{(j)} = \int dE P^{(j)}_{\infty}(E) \simeq \frac{\Gamma_j}{\Gamma}.
\end{equation}
If the decay channels are represented by states $\ket{j}\otimes\ket{\omega}$ in which a photon of frequency $\omega$ is emitted by a bound system in the transition from an initial state $\ket{i}$ of energy $\hbar\omega_i$ towards a specific final state $\ket{j}$ of energy $\hbar\omega_j$, the spectral distribution of final states can be conveniently represented in terms of the photon frequency,
\begin{align}
S^{(j)} (\omega) & = \hbar P_{\infty}^{(j)}(\hbar(\omega_j + \omega)) \nonumber \\ & \simeq \frac{1}{2\pi} \frac{ \Gamma_j }{(\omega - (\omega_{i} - \omega_j + \Delta))^2 + \frac{\Gamma^2}{4}} ,
\end{align}
which corresponds to the quantity in Eq.~\eqref{distribution}.

\bibliography{asym2}

\begin{thebibliography}{23}%
\makeatletter
\providecommand \@ifxundefined [1]{%
 \@ifx{#1\undefined}
}%
\providecommand \@ifnum [1]{%
 \ifnum #1\expandafter \@firstoftwo
 \else \expandafter \@secondoftwo
 \fi
}%
\providecommand \@ifx [1]{%
 \ifx #1\expandafter \@firstoftwo
 \else \expandafter \@secondoftwo
 \fi
}%
\providecommand \natexlab [1]{#1}%
\providecommand \enquote  [1]{``#1''}%
\providecommand \bibnamefont  [1]{#1}%
\providecommand \bibfnamefont [1]{#1}%
\providecommand \citenamefont [1]{#1}%
\providecommand \href@noop [0]{\@secondoftwo}%
\providecommand \href [0]{\begingroup \@sanitize@url \@href}%
\providecommand \@href[1]{\@@startlink{#1}\@@href}%
\providecommand \@@href[1]{\endgroup#1\@@endlink}%
\providecommand \@sanitize@url [0]{\catcode `\\12\catcode `\$12\catcode
  `\&12\catcode `\#12\catcode `\^12\catcode `\_12\catcode `\%12\relax}%
\providecommand \@@startlink[1]{}%
\providecommand \@@endlink[0]{}%
\providecommand \url  [0]{\begingroup\@sanitize@url \@url }%
\providecommand \@url [1]{\endgroup\@href {#1}{\urlprefix }}%
\providecommand \urlprefix  [0]{URL }%
\providecommand \Eprint [0]{\href }%
\providecommand \doibase [0]{https://doi.org/}%
\providecommand \selectlanguage [0]{\@gobble}%
\providecommand \bibinfo  [0]{\@secondoftwo}%
\providecommand \bibfield  [0]{\@secondoftwo}%
\providecommand \translation [1]{[#1]}%
\providecommand \BibitemOpen [0]{}%
\providecommand \bibitemStop [0]{}%
\providecommand \bibitemNoStop [0]{.\EOS\space}%
\providecommand \EOS [0]{\spacefactor3000\relax}%
\providecommand \BibitemShut  [1]{\csname bibitem#1\endcsname}%
\let\auto@bib@innerbib\@empty
\bibitem [{\citenamefont {Xie}\ \emph {et~al.}(2017)\citenamefont {Xie},
  \citenamefont {Zhong}, \citenamefont {Batchelor},\ and\ \citenamefont
  {Lee}}]{xie2017}%
  \BibitemOpen
  \bibfield  {author} {\bibinfo {author} {\bibfnamefont {Q.}~\bibnamefont
  {Xie}}, \bibinfo {author} {\bibfnamefont {H.}~\bibnamefont {Zhong}}, \bibinfo
  {author} {\bibfnamefont {M.~T.}\ \bibnamefont {Batchelor}},\ and\ \bibinfo
  {author} {\bibfnamefont {C.}~\bibnamefont {Lee}},\ }\bibfield  {title}
  {\bibinfo {title} {The quantum rabi model: solution and dynamics},\
  }\href@noop {} {\bibfield  {journal} {\bibinfo  {journal} {Journal of Physics
  A: Mathematical and Theoretical}\ }\textbf {\bibinfo {volume} {50}},\
  \bibinfo {pages} {113001} (\bibinfo {year} {2017})}\BibitemShut {NoStop}%
\bibitem [{\citenamefont {Shore}\ and\ \citenamefont
  {Knight}(1993)}]{shore1993}%
  \BibitemOpen
  \bibfield  {author} {\bibinfo {author} {\bibfnamefont {B.~W.}\ \bibnamefont
  {Shore}}\ and\ \bibinfo {author} {\bibfnamefont {P.~L.}\ \bibnamefont
  {Knight}},\ }\bibfield  {title} {\bibinfo {title} {The jaynes-cummings
  model},\ }\href@noop {} {\bibfield  {journal} {\bibinfo  {journal} {Journal
  of Modern Optics}\ }\textbf {\bibinfo {volume} {40}},\ \bibinfo {pages}
  {1195} (\bibinfo {year} {1993})}\BibitemShut {NoStop}%
\bibitem [{\citenamefont {Forn-D{\'\i}az}\ \emph {et~al.}(2019)\citenamefont
  {Forn-D{\'\i}az}, \citenamefont {Lamata}, \citenamefont {Rico}, \citenamefont
  {Kono},\ and\ \citenamefont {Solano}}]{forn2019}%
  \BibitemOpen
  \bibfield  {author} {\bibinfo {author} {\bibfnamefont {P.}~\bibnamefont
  {Forn-D{\'\i}az}}, \bibinfo {author} {\bibfnamefont {L.}~\bibnamefont
  {Lamata}}, \bibinfo {author} {\bibfnamefont {E.}~\bibnamefont {Rico}},
  \bibinfo {author} {\bibfnamefont {J.}~\bibnamefont {Kono}},\ and\ \bibinfo
  {author} {\bibfnamefont {E.}~\bibnamefont {Solano}},\ }\bibfield  {title}
  {\bibinfo {title} {Ultrastrong coupling regimes of light-matter
  interaction},\ }\href@noop {} {\bibfield  {journal} {\bibinfo  {journal}
  {Reviews of Modern Physics}\ }\textbf {\bibinfo {volume} {91}},\ \bibinfo
  {pages} {025005} (\bibinfo {year} {2019})}\BibitemShut {NoStop}%
\bibitem [{\citenamefont {Bourassa}\ \emph {et~al.}(2009)\citenamefont
  {Bourassa}, \citenamefont {Gambetta}, \citenamefont {Abdumalikov~Jr},
  \citenamefont {Astafiev}, \citenamefont {Nakamura},\ and\ \citenamefont
  {Blais}}]{bourassa2009}%
  \BibitemOpen
  \bibfield  {author} {\bibinfo {author} {\bibfnamefont {J.}~\bibnamefont
  {Bourassa}}, \bibinfo {author} {\bibfnamefont {J.~M.}\ \bibnamefont
  {Gambetta}}, \bibinfo {author} {\bibfnamefont {A.~A.}\ \bibnamefont
  {Abdumalikov~Jr}}, \bibinfo {author} {\bibfnamefont {O.}~\bibnamefont
  {Astafiev}}, \bibinfo {author} {\bibfnamefont {Y.}~\bibnamefont {Nakamura}},\
  and\ \bibinfo {author} {\bibfnamefont {A.}~\bibnamefont {Blais}},\ }\bibfield
   {title} {\bibinfo {title} {Ultrastrong coupling regime of cavity qed with
  phase-biased flux qubits},\ }\href@noop {} {\bibfield  {journal} {\bibinfo
  {journal} {Physical Review A}\ }\textbf {\bibinfo {volume} {80}},\ \bibinfo
  {pages} {032109} (\bibinfo {year} {2009})}\BibitemShut {NoStop}%
\bibitem [{\citenamefont {Niemczyk}\ \emph {et~al.}(2010)\citenamefont
  {Niemczyk}, \citenamefont {Deppe}, \citenamefont {Huebl}, \citenamefont
  {Menzel}, \citenamefont {Hocke}, \citenamefont {Schwarz}, \citenamefont
  {Garcia-Ripoll}, \citenamefont {Zueco}, \citenamefont {H{\"u}mmer},
  \citenamefont {Solano} \emph {et~al.}}]{niemczyk2010}%
  \BibitemOpen
  \bibfield  {author} {\bibinfo {author} {\bibfnamefont {T.}~\bibnamefont
  {Niemczyk}}, \bibinfo {author} {\bibfnamefont {F.}~\bibnamefont {Deppe}},
  \bibinfo {author} {\bibfnamefont {H.}~\bibnamefont {Huebl}}, \bibinfo
  {author} {\bibfnamefont {E.}~\bibnamefont {Menzel}}, \bibinfo {author}
  {\bibfnamefont {F.}~\bibnamefont {Hocke}}, \bibinfo {author} {\bibfnamefont
  {M.}~\bibnamefont {Schwarz}}, \bibinfo {author} {\bibfnamefont
  {J.}~\bibnamefont {Garcia-Ripoll}}, \bibinfo {author} {\bibfnamefont
  {D.}~\bibnamefont {Zueco}}, \bibinfo {author} {\bibfnamefont
  {T.}~\bibnamefont {H{\"u}mmer}}, \bibinfo {author} {\bibfnamefont
  {E.}~\bibnamefont {Solano}}, \emph {et~al.},\ }\bibfield  {title} {\bibinfo
  {title} {Circuit quantum electrodynamics in the ultrastrong-coupling
  regime},\ }\href@noop {} {\bibfield  {journal} {\bibinfo  {journal} {Nature
  Physics}\ }\textbf {\bibinfo {volume} {6}},\ \bibinfo {pages} {772} (\bibinfo
  {year} {2010})}\BibitemShut {NoStop}%
\bibitem [{\citenamefont {G{\"u}nter}\ \emph {et~al.}(2009)\citenamefont
  {G{\"u}nter}, \citenamefont {Anappara}, \citenamefont {Hees}, \citenamefont
  {Sell}, \citenamefont {Biasiol}, \citenamefont {Sorba}, \citenamefont
  {De~Liberato}, \citenamefont {Ciuti}, \citenamefont {Tredicucci},
  \citenamefont {Leitenstorfer} \emph {et~al.}}]{gunter2009}%
  \BibitemOpen
  \bibfield  {author} {\bibinfo {author} {\bibfnamefont {G.}~\bibnamefont
  {G{\"u}nter}}, \bibinfo {author} {\bibfnamefont {A.~A.}\ \bibnamefont
  {Anappara}}, \bibinfo {author} {\bibfnamefont {J.}~\bibnamefont {Hees}},
  \bibinfo {author} {\bibfnamefont {A.}~\bibnamefont {Sell}}, \bibinfo {author}
  {\bibfnamefont {G.}~\bibnamefont {Biasiol}}, \bibinfo {author} {\bibfnamefont
  {L.}~\bibnamefont {Sorba}}, \bibinfo {author} {\bibfnamefont
  {S.}~\bibnamefont {De~Liberato}}, \bibinfo {author} {\bibfnamefont
  {C.}~\bibnamefont {Ciuti}}, \bibinfo {author} {\bibfnamefont
  {A.}~\bibnamefont {Tredicucci}}, \bibinfo {author} {\bibfnamefont
  {A.}~\bibnamefont {Leitenstorfer}}, \emph {et~al.},\ }\bibfield  {title}
  {\bibinfo {title} {Sub-cycle switch-on of ultrastrong light--matter
  interaction},\ }\href@noop {} {\bibfield  {journal} {\bibinfo  {journal}
  {Nature}\ }\textbf {\bibinfo {volume} {458}},\ \bibinfo {pages} {178}
  (\bibinfo {year} {2009})}\BibitemShut {NoStop}%
\bibitem [{\citenamefont {Zhang}\ \emph {et~al.}(2016)\citenamefont {Zhang},
  \citenamefont {Lou}, \citenamefont {Li}, \citenamefont {Reno}, \citenamefont
  {Pan}, \citenamefont {Watson}, \citenamefont {Manfra},\ and\ \citenamefont
  {Kono}}]{zhang2016}%
  \BibitemOpen
  \bibfield  {author} {\bibinfo {author} {\bibfnamefont {Q.}~\bibnamefont
  {Zhang}}, \bibinfo {author} {\bibfnamefont {M.}~\bibnamefont {Lou}}, \bibinfo
  {author} {\bibfnamefont {X.}~\bibnamefont {Li}}, \bibinfo {author}
  {\bibfnamefont {J.~L.}\ \bibnamefont {Reno}}, \bibinfo {author}
  {\bibfnamefont {W.}~\bibnamefont {Pan}}, \bibinfo {author} {\bibfnamefont
  {J.~D.}\ \bibnamefont {Watson}}, \bibinfo {author} {\bibfnamefont {M.~J.}\
  \bibnamefont {Manfra}},\ and\ \bibinfo {author} {\bibfnamefont
  {J.}~\bibnamefont {Kono}},\ }\bibfield  {title} {\bibinfo {title} {Collective
  non-perturbative coupling of 2d electrons with high-quality-factor terahertz
  cavity photons},\ }\href@noop {} {\bibfield  {journal} {\bibinfo  {journal}
  {Nature Physics}\ }\textbf {\bibinfo {volume} {12}},\ \bibinfo {pages} {1005}
  (\bibinfo {year} {2016})}\BibitemShut {NoStop}%
\bibitem [{\citenamefont {Crespi}\ \emph {et~al.}(2012)\citenamefont {Crespi},
  \citenamefont {Longhi},\ and\ \citenamefont {Osellame}}]{crespi2012}%
  \BibitemOpen
  \bibfield  {author} {\bibinfo {author} {\bibfnamefont {A.}~\bibnamefont
  {Crespi}}, \bibinfo {author} {\bibfnamefont {S.}~\bibnamefont {Longhi}},\
  and\ \bibinfo {author} {\bibfnamefont {R.}~\bibnamefont {Osellame}},\
  }\bibfield  {title} {\bibinfo {title} {Photonic realization of the quantum
  rabi model},\ }\href@noop {} {\bibfield  {journal} {\bibinfo  {journal}
  {Physical review letters}\ }\textbf {\bibinfo {volume} {108}},\ \bibinfo
  {pages} {163601} (\bibinfo {year} {2012})}\BibitemShut {NoStop}%
\bibitem [{\citenamefont {George}\ \emph {et~al.}(2016)\citenamefont {George},
  \citenamefont {Chervy}, \citenamefont {Shalabney}, \citenamefont {Devaux},
  \citenamefont {Hiura}, \citenamefont {Genet},\ and\ \citenamefont
  {Ebbesen}}]{george2016}%
  \BibitemOpen
  \bibfield  {author} {\bibinfo {author} {\bibfnamefont {J.}~\bibnamefont
  {George}}, \bibinfo {author} {\bibfnamefont {T.}~\bibnamefont {Chervy}},
  \bibinfo {author} {\bibfnamefont {A.}~\bibnamefont {Shalabney}}, \bibinfo
  {author} {\bibfnamefont {E.}~\bibnamefont {Devaux}}, \bibinfo {author}
  {\bibfnamefont {H.}~\bibnamefont {Hiura}}, \bibinfo {author} {\bibfnamefont
  {C.}~\bibnamefont {Genet}},\ and\ \bibinfo {author} {\bibfnamefont {T.~W.}\
  \bibnamefont {Ebbesen}},\ }\bibfield  {title} {\bibinfo {title} {Multiple
  rabi splittings under ultrastrong vibrational coupling},\ }\href@noop {}
  {\bibfield  {journal} {\bibinfo  {journal} {Physical review letters}\
  }\textbf {\bibinfo {volume} {117}},\ \bibinfo {pages} {153601} (\bibinfo
  {year} {2016})}\BibitemShut {NoStop}%
\bibitem [{\citenamefont {Schneeweiss}\ \emph {et~al.}(2018)\citenamefont
  {Schneeweiss}, \citenamefont {Dareau},\ and\ \citenamefont
  {Sayrin}}]{schneeweiss2018}%
  \BibitemOpen
  \bibfield  {author} {\bibinfo {author} {\bibfnamefont {P.}~\bibnamefont
  {Schneeweiss}}, \bibinfo {author} {\bibfnamefont {A.}~\bibnamefont
  {Dareau}},\ and\ \bibinfo {author} {\bibfnamefont {C.}~\bibnamefont
  {Sayrin}},\ }\bibfield  {title} {\bibinfo {title} {Cold-atom-based
  implementation of the quantum rabi model},\ }\href@noop {} {\bibfield
  {journal} {\bibinfo  {journal} {Physical Review A}\ }\textbf {\bibinfo
  {volume} {98}},\ \bibinfo {pages} {021801} (\bibinfo {year}
  {2018})}\BibitemShut {NoStop}%
\bibitem [{\citenamefont {Braak}(2011)}]{braak2011}%
  \BibitemOpen
  \bibfield  {author} {\bibinfo {author} {\bibfnamefont {D.}~\bibnamefont
  {Braak}},\ }\bibfield  {title} {\bibinfo {title} {Integrability of the rabi
  model},\ }\href@noop {} {\bibfield  {journal} {\bibinfo  {journal} {Physical
  Review Letters}\ }\textbf {\bibinfo {volume} {107}},\ \bibinfo {pages}
  {100401} (\bibinfo {year} {2011})}\BibitemShut {NoStop}%
\bibitem [{\citenamefont {Chen}\ \emph {et~al.}(2012)\citenamefont {Chen},
  \citenamefont {Wang}, \citenamefont {He}, \citenamefont {Liu},\ and\
  \citenamefont {Wang}}]{chen2012}%
  \BibitemOpen
  \bibfield  {author} {\bibinfo {author} {\bibfnamefont {Q.-H.}\ \bibnamefont
  {Chen}}, \bibinfo {author} {\bibfnamefont {C.}~\bibnamefont {Wang}}, \bibinfo
  {author} {\bibfnamefont {S.}~\bibnamefont {He}}, \bibinfo {author}
  {\bibfnamefont {T.}~\bibnamefont {Liu}},\ and\ \bibinfo {author}
  {\bibfnamefont {K.-L.}\ \bibnamefont {Wang}},\ }\bibfield  {title} {\bibinfo
  {title} {Exact solvability of the quantum rabi model using bogoliubov
  operators},\ }\href@noop {} {\bibfield  {journal} {\bibinfo  {journal}
  {Physical Review A}\ }\textbf {\bibinfo {volume} {86}},\ \bibinfo {pages}
  {023822} (\bibinfo {year} {2012})}\BibitemShut {NoStop}%
\bibitem [{\citenamefont {Kibis}\ \emph {et~al.}(2009)\citenamefont {Kibis},
  \citenamefont {Slepyan}, \citenamefont {Maksimenko},\ and\ \citenamefont
  {Hoffmann}}]{kibis2009}%
  \BibitemOpen
  \bibfield  {author} {\bibinfo {author} {\bibfnamefont {O.}~\bibnamefont
  {Kibis}}, \bibinfo {author} {\bibfnamefont {G.~Y.}\ \bibnamefont {Slepyan}},
  \bibinfo {author} {\bibfnamefont {S.}~\bibnamefont {Maksimenko}},\ and\
  \bibinfo {author} {\bibfnamefont {A.}~\bibnamefont {Hoffmann}},\ }\bibfield
  {title} {\bibinfo {title} {Matter coupling to strong electromagnetic fields
  in two-level quantum systems with broken inversion symmetry},\ }\href@noop {}
  {\bibfield  {journal} {\bibinfo  {journal} {Physical review letters}\
  }\textbf {\bibinfo {volume} {102}},\ \bibinfo {pages} {023601} (\bibinfo
  {year} {2009})}\BibitemShut {NoStop}%
\bibitem [{\citenamefont {Paspalakis}\ \emph {et~al.}(2013)\citenamefont
  {Paspalakis}, \citenamefont {Boviatsis},\ and\ \citenamefont
  {Baskoutas}}]{paspalakis2013}%
  \BibitemOpen
  \bibfield  {author} {\bibinfo {author} {\bibfnamefont {E.}~\bibnamefont
  {Paspalakis}}, \bibinfo {author} {\bibfnamefont {J.}~\bibnamefont
  {Boviatsis}},\ and\ \bibinfo {author} {\bibfnamefont {S.}~\bibnamefont
  {Baskoutas}},\ }\bibfield  {title} {\bibinfo {title} {Effects of probe field
  intensity in nonlinear optical processes in asymmetric semiconductor quantum
  dots},\ }\href@noop {} {\bibfield  {journal} {\bibinfo  {journal} {Journal of
  Applied Physics}\ }\textbf {\bibinfo {volume} {114}},\ \bibinfo {pages}
  {153107} (\bibinfo {year} {2013})}\BibitemShut {NoStop}%
\bibitem [{\citenamefont {Chestnov}\ \emph {et~al.}(2017)\citenamefont
  {Chestnov}, \citenamefont {Shahnazaryan}, \citenamefont {Alodjants},\ and\
  \citenamefont {Shelykh}}]{chestnov2017}%
  \BibitemOpen
  \bibfield  {author} {\bibinfo {author} {\bibfnamefont {I.~Y.}\ \bibnamefont
  {Chestnov}}, \bibinfo {author} {\bibfnamefont {V.~A.}\ \bibnamefont
  {Shahnazaryan}}, \bibinfo {author} {\bibfnamefont {A.~P.}\ \bibnamefont
  {Alodjants}},\ and\ \bibinfo {author} {\bibfnamefont {I.~A.}\ \bibnamefont
  {Shelykh}},\ }\bibfield  {title} {\bibinfo {title} {Terahertz lasing in
  ensemble of asymmetric quantum dots},\ }\href@noop {} {\bibfield  {journal}
  {\bibinfo  {journal} {Acs Photonics}\ }\textbf {\bibinfo {volume} {4}},\
  \bibinfo {pages} {2726} (\bibinfo {year} {2017})}\BibitemShut {NoStop}%
\bibitem [{\citenamefont {G{\l}adysz}\ \emph {et~al.}(2020)\citenamefont
  {G{\l}adysz}, \citenamefont {Wcis{\l}o},\ and\ \citenamefont
  {S{\l}owik}}]{gladysz2020}%
  \BibitemOpen
  \bibfield  {author} {\bibinfo {author} {\bibfnamefont {P.}~\bibnamefont
  {G{\l}adysz}}, \bibinfo {author} {\bibfnamefont {P.}~\bibnamefont
  {Wcis{\l}o}},\ and\ \bibinfo {author} {\bibfnamefont {K.}~\bibnamefont
  {S{\l}owik}},\ }\bibfield  {title} {\bibinfo {title} {Propagation of
  optically tunable coherent radiation in a gas of polar molecules},\
  }\href@noop {} {\bibfield  {journal} {\bibinfo  {journal} {Scientific
  reports}\ }\textbf {\bibinfo {volume} {10}},\ \bibinfo {pages} {1} (\bibinfo
  {year} {2020})}\BibitemShut {NoStop}%
\bibitem [{\citenamefont {Koppenh{\"o}fer}\ and\ \citenamefont
  {Marthaler}(2016)}]{koppenhofer2016}%
  \BibitemOpen
  \bibfield  {author} {\bibinfo {author} {\bibfnamefont {M.}~\bibnamefont
  {Koppenh{\"o}fer}}\ and\ \bibinfo {author} {\bibfnamefont {M.}~\bibnamefont
  {Marthaler}},\ }\bibfield  {title} {\bibinfo {title} {Creation of a squeezed
  photon distribution using artificial atoms with broken inversion symmetry},\
  }\href@noop {} {\bibfield  {journal} {\bibinfo  {journal} {Physical Review
  A}\ }\textbf {\bibinfo {volume} {93}},\ \bibinfo {pages} {023831} (\bibinfo
  {year} {2016})}\BibitemShut {NoStop}%
\bibitem [{\citenamefont {Ant{\'o}n}\ \emph {et~al.}(2017)\citenamefont
  {Ant{\'o}n}, \citenamefont {Maede-Razavi}, \citenamefont {Carreno},
  \citenamefont {Thanopulos},\ and\ \citenamefont {Paspalakis}}]{anton2017}%
  \BibitemOpen
  \bibfield  {author} {\bibinfo {author} {\bibfnamefont {M.}~\bibnamefont
  {Ant{\'o}n}}, \bibinfo {author} {\bibfnamefont {S.}~\bibnamefont
  {Maede-Razavi}}, \bibinfo {author} {\bibfnamefont {F.}~\bibnamefont
  {Carreno}}, \bibinfo {author} {\bibfnamefont {I.}~\bibnamefont
  {Thanopulos}},\ and\ \bibinfo {author} {\bibfnamefont {E.}~\bibnamefont
  {Paspalakis}},\ }\bibfield  {title} {\bibinfo {title} {Optical and microwave
  control of resonance fluorescence and squeezing spectra in a polar
  molecule},\ }\href@noop {} {\bibfield  {journal} {\bibinfo  {journal}
  {Physical Review A}\ }\textbf {\bibinfo {volume} {96}},\ \bibinfo {pages}
  {063812} (\bibinfo {year} {2017})}\BibitemShut {NoStop}%
\bibitem [{\citenamefont {Scala}\ \emph {et~al.}(2020)\citenamefont {Scala},
  \citenamefont {Pepe}, \citenamefont {Facchi}, \citenamefont {Pascazio},\ and\
  \citenamefont {S\l{}owik}}]{scala2020}%
  \BibitemOpen
  \bibfield  {author} {\bibinfo {author} {\bibfnamefont {G.}~\bibnamefont
  {Scala}}, \bibinfo {author} {\bibfnamefont {F.~V.}\ \bibnamefont {Pepe}},
  \bibinfo {author} {\bibfnamefont {P.}~\bibnamefont {Facchi}}, \bibinfo
  {author} {\bibfnamefont {S.}~\bibnamefont {Pascazio}},\ and\ \bibinfo
  {author} {\bibfnamefont {K.}~\bibnamefont {S\l{}owik}},\ }\bibfield  {title}
  {\bibinfo {title} {Light interaction with extended quantum systems in
  dispersive media},\ }\bibfield  {journal} {\bibinfo  {journal} {New Journal
  of Physics}\ }\href {https://doi.org/10.1088/1367-2630/abd204}
  {10.1088/1367-2630/abd204} (\bibinfo {year} {2020})\BibitemShut {NoStop}%
\bibitem [{\citenamefont {Savenko}\ \emph {et~al.}(2012)\citenamefont
  {Savenko}, \citenamefont {Kibis},\ and\ \citenamefont
  {Shelykh}}]{savenko2012}%
  \BibitemOpen
  \bibfield  {author} {\bibinfo {author} {\bibfnamefont {I.}~\bibnamefont
  {Savenko}}, \bibinfo {author} {\bibfnamefont {O.}~\bibnamefont {Kibis}},\
  and\ \bibinfo {author} {\bibfnamefont {I.~A.}\ \bibnamefont {Shelykh}},\
  }\bibfield  {title} {\bibinfo {title} {Asymmetric quantum dot in a
  microcavity as a nonlinear optical element},\ }\href@noop {} {\bibfield
  {journal} {\bibinfo  {journal} {Physical Review A}\ }\textbf {\bibinfo
  {volume} {85}},\ \bibinfo {pages} {053818} (\bibinfo {year}
  {2012})}\BibitemShut {NoStop}%
\bibitem [{\citenamefont {Lu}\ \emph {et~al.}(2018)\citenamefont {Lu},
  \citenamefont {Chen}, \citenamefont {Zou},\ and\ \citenamefont
  {Xie}}]{lu2018}%
  \BibitemOpen
  \bibfield  {author} {\bibinfo {author} {\bibfnamefont {Q.}~\bibnamefont
  {Lu}}, \bibinfo {author} {\bibfnamefont {X.}~\bibnamefont {Chen}}, \bibinfo
  {author} {\bibfnamefont {C.-L.}\ \bibnamefont {Zou}},\ and\ \bibinfo {author}
  {\bibfnamefont {S.}~\bibnamefont {Xie}},\ }\bibfield  {title} {\bibinfo
  {title} {Extreme terahertz electric-field enhancement in high-q photonic
  crystal slab cavity with nanoholes},\ }\href@noop {} {\bibfield  {journal}
  {\bibinfo  {journal} {Optics Express}\ }\textbf {\bibinfo {volume} {26}},\
  \bibinfo {pages} {30851} (\bibinfo {year} {2018})}\BibitemShut {NoStop}%
\bibitem [{\citenamefont {Vogt}\ and\ \citenamefont
  {Leonhardt}(2018)}]{vogt2018}%
  \BibitemOpen
  \bibfield  {author} {\bibinfo {author} {\bibfnamefont {D.~W.}\ \bibnamefont
  {Vogt}}\ and\ \bibinfo {author} {\bibfnamefont {R.}~\bibnamefont
  {Leonhardt}},\ }\bibfield  {title} {\bibinfo {title} {Ultra-high q terahertz
  whispering-gallery modes in a silicon resonator},\ }\href@noop {} {\bibfield
  {journal} {\bibinfo  {journal} {APL Photonics}\ }\textbf {\bibinfo {volume}
  {3}},\ \bibinfo {pages} {051702} (\bibinfo {year} {2018})}\BibitemShut
  {NoStop}%
\bibitem [{\citenamefont {Zografopoulos}\ and\ \citenamefont
  {Beccherelli}(2015)}]{zografopoulos2015}%
  \BibitemOpen
  \bibfield  {author} {\bibinfo {author} {\bibfnamefont {D.~C.}\ \bibnamefont
  {Zografopoulos}}\ and\ \bibinfo {author} {\bibfnamefont {R.}~\bibnamefont
  {Beccherelli}},\ }\bibfield  {title} {\bibinfo {title} {Tunable terahertz
  fishnet metamaterials based on thin nematic liquid crystal layers for fast
  switching},\ }\href@noop {} {\bibfield  {journal} {\bibinfo  {journal}
  {Scientific Reports}\ }\textbf {\bibinfo {volume} {5}},\ \bibinfo {pages}
  {13137} (\bibinfo {year} {2015})}\BibitemShut {NoStop}%
\end{thebibliography}%

\end{document}